\documentclass[
10pt,
showpacs,preprintnumbers,footinbib,
amsmath,amssymb,
aps,
prc,twocolumn,groupedaddress,superscriptaddress,
showkeys
]{revtex4-1}
\usepackage{graphicx}
\usepackage{dcolumn}
\usepackage{bm}
\usepackage[colorlinks=true,urlcolor=blue,citecolor=blue]{hyperref}
\usepackage{color}
\usepackage{lipsum}

\begin{document}

\title{Microscopic calculations of nuclear level densities with the Lanczos
method }

\email{ormand1@llnl.gov}
\author{W. E. Ormand}
\affiliation{Lawrence Livermore National Laboratory, P.O. Box 808, L-414,
Livermore, California 94551, USA}
\affiliation{Department of Physics and the National Superconducting Cyclotron
Laboratory, \\
   Michigan State University, East Lansing, MI 42284-1321, USA}
\author{B. A. Brown}
\affiliation{Department of Physics and the National Superconducting Cyclotron
Laboratory, \\
   Michigan State University, East Lansing, MI 42284-1321, USA}

\date{\today}

\begin{abstract}
A new method for computing the density of states in nuclei making use of an extrapolated form of the tri-diagonal matrix obtained from the Lanczos method is presented.  It will be shown that the global, average properties of the entire Lanczos matrix can be  predicted from just four Lanczos iterations.  The extrapolated Lanczos matrix (ELM) approach provides for an accurate computation of the density of states described within the configuration space, which, in some cases, is sufficient to accurately calculate the density of states at, or near, the neutron separation energy. Comparisons between theory and experiment are shown for $^{57}$Fe, $^{74}$Ge, and $^{76}$Ge. In addition, we show results for the $J$-dependence of moments and the level density for these three nuclei.

\end{abstract}


\maketitle

\section{Introduction}
The density of states is a fundamental property of nuclear structure and plays a key 
role in nuclear reactions.  An important example is the radiative capture of neutrons on
short-lived nuclei, which, through the r-process \cite{r-process} in supernovae and/or
neutron-star mergers \cite{merg}, are thought to be responsible for the synthesis of the 
elements heavier than iron. Ideally, these reactions can be measured or
constrained by experiment. Unfortunately, in most cases, the target nuclei
are so short lived that direct measurement is not possible, and the only
alternative is to rely on theoretical calculations or indirect
measurements such as surrogates \cite{surr}, which are themselves reliant on theoretical 
input. 

Nuclear reaction approaches such as Hauser-Feshbach \cite{Hauser-Feshbach} can give an accurate description of the neutron-capture cross section. However, the Hauser-Feshbach model requires accurate knowledge of the density of states up to the neutron-decay threshold. A challenge in nuclear theory is to accurately compute the density of states. This is difficult because of the sheer number of levels and configurations and the strong nature of the nuclear Hamiltonian. One microscopic  approach is to account for correlations at the Hartree-Fock level and to ``count" non-interacting levels within the corresponding mean-field single-particle space \cite{Goriely}. Another is to use the Shell-Model Monte Carlo (SMMC)~\cite{AFMC,AFMC-2}, which utilizes auxiliary fields to compute the thermal trace for the energy, from which, the density of states can be extracted from the inverse Laplace transform of the partition function \cite{AFMC-rho}. A limitation of the SMMC is the sign problem, which primarily limits the approach to schematic interactions \cite{AFMC-2}.  Moments methods, derived from random matrix theory and statistical spectroscopy, can be used to construct spin and parity dependent level densities for realistic Hamiltonians~\cite{Mon75,mom,Horoi}. Moments method, however, have been limited by the ability to compute higher moments of the Hamiltonian, the overall structural form density of states, and must be matched to the exact energies for low-lying states. The stochastic estimation method \cite{shi}  has a computational cost that is almost the same order as the Lanczos method used here and requires a special computer code to apply the shifted Krylov-subspace method \cite{26,27}.
 
In this article, we report on a new framework to provide an accurate description of the statistical properties of a model Hamiltonian. Our specific application is the calculation of the nuclear density of states within the configuration-interaction approach using fully realistic nuclear Hamiltonians. From universal properties of the Lanczos algorithm,  we will demonstrate that the first eight moments of the Hamiltonian can be obtained from just four Lanczos iterations, which, in turn, can provide an accurate description of the averaged, or global, properties of the nuclear system within the defined Hilbert space. Several procedures to extract the density of states for model Hamiltonians are presented here: 1) extrapolating the tri-diagonal Lanczos matrix well beyond what is computationally viable, leading to an extrapolated Lanczos method (ELM) to efficiently compute  compute the density of states within the  configuration-interaction method; 2) an analytic continuation of the ELM method; and 3) an approximation of the level density based on the binomial distribution.

\section{Nuclear Structure Model}
The principal goal behind nuclear-structure models is to find energy eigenvalues and wave functions for the nuclear Hamiltonian within a well-defined Hilbert space. In the nuclear shell model \cite{shell-model}, or configuration interaction, the Hilbert space is defined by a set of  orbits, usually denoted by the principal quantum number $n$, orbital angular momentum $l$, and angular momentum $j$. The nuclear wave functions are constructed through a set of basis states obtained by filling these orbits following the Pauli principle. The basis states can consist of a set of Slater determinants with well defined $z$-projection of angular momentum, $J_z=M$, in the so-called $M$-scheme, or by projecting angular momentum (and possibly isospin) onto the $M$-scheme Slater determinants. The $N$ many-body basis states, $|\psi_{i}\rangle$, spanning the Hilbert space are used to construct the
full solution, i.e., $ | \Psi\rangle=\sum_{i} c_{i}|\psi_{i}\rangle  $. The coefficients $  c_{i}  $ are found
by computing the matrix elements of the Hamiltonian, $  H_{ij}=\langle \psi_{i}| \hat H | \psi_{j}\rangle  $, and
diagonalizing the resulting Hermitian matrix. One of the most effective methods
to find the lowest eigenvalues is the Lanczos algorithm \cite{Lanczos}, which starts with
an arbitrary vector $| v_{1} \rangle $ in the Hilbert space, and through successive
operations of $  \hat{H}  $, the matrix $\textbf{H}$ is transformed into 
tri-diagonal form. The
first three terms are
\begin{align}
\hat{H}| v_{1}\rangle &= \alpha_{1} | v_{1} \rangle +\beta_{1}| v_{2}\rangle,  \nonumber \\ 
\hat{H}| v_{2}\rangle &= \beta_{1}| v_{1}\rangle+ \alpha_{2}| v_{2} \rangle + \beta_{2}| v_{3}\rangle, \nonumber \\
\hat{H}| v_{3}\rangle &= \hspace{1.375cm}\beta_{2}| v_{2}\rangle+\alpha_{3}| v_{3}\rangle + \beta_{3} | v_{4}\rangle,     
\end{align}
and the $|v_i\rangle$ form an orthonormal set. 
In practice this amounts to applying $\hat H$ to the Lanczos vectors, and extracting the matrix 
elements through subsequent dot-product operations and reorthogonalization, e.g., 
$\alpha_1 = \langle v_1 | \hat H | v_1 \rangle$, and 
$\beta_1^2 = \langle v_1 | (\hat H^\dagger - \alpha_1) (\hat H - \alpha_1) | v_1\rangle$ 
(note that the phase of any of the $\beta_i$ is arbitrary).
The power of the Lanczos algorithm is that
following successive applications of $  \hat{H}  $ (iterations), the eigenvalues of the
tri-diagonal matrix quickly converge to the extreme eigenvalues of the full
matrix. Typically, the lowest energy in the model space, $E_0$, is
obtained in approximately 30 iterations regardless the matrix dimension. 

Of particular interest is the behavior of the tri-diagonal matrix elements with 
increasing iterations. After several iterations, the diagonal elements, $  \alpha_{i}  $, are
roughly constant and nearly equal to the first moment
$  H_1 = \frac{1}{N} {\rm {Tr}}[\hat H] = \frac{1}{N} \sum_{i} H_{ii}$. At the same time, 
the off-diagonal elements, $
\beta_i$, generally decrease to zero as $i \rightarrow N$, and exhibit a 
Gaussian-like behavior~\cite{zuker}. 

In this work, we will examine the level density for selected Cr, Fe, and Ge isotopes within the framework of the nuclear shell model. All shell-model calculations were performed using angular momentum projected basis states with the NuShellX shell-model code~\cite{nushellx} framework. For the Fe isotopes, the model space is comprised of the $0f_{7/2}$, $0f_{5/2}$, $1p_{3/2}$, and $1p_{1/2}$ orbitals and the Hamiltonian is defined by the one- and two-body matrix elements of the GXPF1A interaction of Ref.~\cite{gxpf1a}. The model space for the Ge isotopes consists of the $0f_{5/2}$, $1p_{3/2}$, $1p_{1/2}$, $0g_{9/2}$ orbitals. For the Ge isotopes, we present results for two different empirical Hamiltonians:  1) $jj44b$ defined in the appendix of Ref.~\cite{Muk} and 2) jun45 of Ref.~\cite{Homna_2009}. Note that there are no spurious center-of-mass excitations in either of these model spaces. 

\section{Computing the Hamiltonian moments with Lanczos}

At its core, the Lanczos algorithm is a really moment method; efficiently computing $ 2n$ moments of $  \hat H  $ with respect to the initial pivot vector $ | v_{1}\rangle  $ after $ n$ iterations. With the choice of $|v_1\rangle = \frac{1} {\sqrt{N}} \sum_i \phi_i | \psi_i\rangle$, where $\phi_i$ is a random phase, we find it is possible to efficiently compute several moments of the Hamiltonian with just a few Lanczos iterations. This is illustrated by the first Lanczos matrix element $\alpha_1$ given by
\begin{equation}
\label{eq:alpha_1}
\alpha_1 = \frac{1}{N}\sum_i H_{ii} + \sum_{i \ne j} \frac{\phi_i \phi_j}{N}H_{ji} 
= H_1 + \sum_{i \ne j} \frac{\phi_i \phi_j}{N}H_{ji}.
\end{equation}
The remainder in Eq. (\ref{eq:alpha_1}) is generally small due to cancellations caused by the random phases and a diminishing magnitude due to the large factor $N$ in the denominator. Thus, for systems with large dimensions $\alpha_1 \approx H_1$. If needed, higher accuracy can be obtained by using different random initial pivots and averaging. A small remainder in Eq.~(\ref{eq:alpha_1}) then suggests a strategy to compute even higher moments $\hat H$ via 
\begin{equation}
M_k  = \frac{1}{N} {\rm{Tr}}[(\hat H - H_1)^k] \approx \langle v_1| (\hat H - \alpha_1)^k | v_1\rangle.
\end{equation}
To compute the moments with Lanczos iterations, we note the recurrence relation for the $n^{th}$ Lanczos vector
\begin{equation}
|v_n \rangle = \frac{\hat h - \alpha_{n-1}+\alpha_1}{\beta_{n-1}}|v_{n-1} \rangle - \frac{\beta_{n-2}}{\beta_{n-1}}|v_{n-2}\rangle,
\end{equation}
with $\hat h = \hat H - \alpha_1$ and $|v_2\rangle = \frac{\hat h}{\beta_1}|v_1\rangle$. In the case that the remainder elements are small, we have the approximation $M_k \approx \langle v_1 | \hat h^k | v_1 \rangle$, which can be extracted from the Lanczos matrix elements through successive application of the recurrence relation, collecting powers of $\hat h$, and back substituting for previous moments. From the $n^{th}$ Lanczos iteration, which gives the Lanczos vectors up to $v_{n+1}$, the moment $M_{n+1}$ can be obtained from the normalization condition $\langle v_{n+1} | v_{n+1}\rangle = 1$, while the moment $M_n$ can be extracted from the orthogonality of the Lanczos vectors, i.e.,  $\langle v_n | v_{n+1}\rangle = 0$.
For example, $M_2$ can be found from normalizing $|v_2\rangle$
\begin{equation}
\langle v_2 | v_2 \rangle =  \frac{ \langle v_1 | {\hat h}^2 | v_1 \rangle}{\beta_1^2} = \frac{M_2}{\beta_1^2} = 1,
\end{equation}
leading to
\begin{equation}
M_2 = \beta_1^2.
\end{equation}
For $M_3$, we use the orthogonality condition
\begin{align}
\langle v_2 | v_3 \rangle &=  \frac{\langle v_2 |\hat h - (\alpha_2-\alpha_1) | v_2\rangle}{\beta_2} - \frac{\beta_1}{\beta_2} \langle v_2|v_1\rangle, \\
 & = \frac{\langle v_1 | \hat h [\hat h - (\alpha_2-\alpha_1)] \hat h| v_1\rangle}{\beta_2\beta_1^2},\\
 & = \frac{M_3 }{\beta_2\beta_1^2} - \frac{\alpha_2-\alpha_1}{\beta_2} = 0,\\
\end{align}
 giving
 \begin{equation}
M_3  = \beta_1^2(-\alpha_1 + \alpha_2).
\end{equation}

Overall, while the derivations are tedious, they are straightforward using the symbolic manipulation program Mathematica. The first eight moments in terms of the matrix elements from the first four Lanczos iterations are given by
\begin{widetext}
\begin{align}
\label{eq:h1}
H_1  = & \alpha_1\\
\label{eq:m2}
M_2  =  &\beta_1^2 \\
\label{eq:m3}
M_3  =  &\beta_1^2(-\alpha_1 + \alpha_2)\\
\label{eq:m4}
M_4  = &\beta_1^2(\alpha_1^2 - 2\alpha_1\alpha_2 + \alpha_2^2 + \beta_1^2 +\beta_2^2) \\
\label{eq:m5}
M_5  = &\beta_1^2\Bigl(-\alpha_1 \left(3 \alpha_2^2+2 \beta_1^2+3 \beta_2^2\right)+\alpha_3
   \beta_2^2 +  2 \alpha_2 \left(\beta_1^2+\beta_2^2\right)-\alpha_1^3+3 \alpha_2 \alpha_1^2+\alpha_2^3\Bigr)\\
\label{eq:m6}
M_6 = &\beta_1^2 \Bigl(3 \alpha_1^2 \left(2 \alpha_2^2+\beta_1^2+2 \beta_2^2\right) - 2 \alpha_1 \left(\alpha_2 \left(3 \beta_1^2+4 \beta_2^2\right) + 2 \alpha_3 \beta_2^2+2 \alpha_2^3\right) + \notag\\
  &\hskip 0.6cm \alpha_3^2 \beta_2^2+2 \alpha_2 \alpha_3 \beta_2^2+3 \alpha_2^2\left(\beta_1^2+\beta_2^2\right)+ \alpha_1^4 -4 \alpha_2 \alpha_1^3+\alpha_2^4  +\beta_1^4+\beta_2^4+2 \beta_1^2 \beta_2^2 + \beta_2^2 \beta_3^2\Bigr)\\
\label{eq:m7}
M_7 =& \beta_1^2 \Bigl(-2 \alpha_1^3 \left(5 \alpha_2^2+2 \beta_1^2+5 \beta_2^2\right) +  
             2 \alpha_1^2 \left(2 \alpha_2 \left(3 \beta_1^2+5 \beta_2^2\right)+5 \alpha_3 \beta_2^2 + 
            5\alpha_2^3\right) - \notag\\
   &\hskip 0.6cm \alpha_1 \Bigl(3 \alpha_2^2 \left(4 \beta_1^2+5 \beta_2^2\right) + 
        10\alpha_3 \alpha_2 \beta_2^2 + 5 \beta_2^2 \left(\alpha_3^2+\beta_2^2+\beta_3^2\right) + 
         5 \alpha_2^4 + 3 \beta_1^4+8 \beta_1^2 \beta_2^2\Bigr) + \notag\\
     &\hskip 0.6 cm 3 \alpha_2^2 \alpha_3 \beta_2^2 + 
         4 \alpha_2^3 \left(\beta_1^2+\beta_2^2\right) + 
       \beta_2^2 \left(2 \alpha_3 \left(\beta_1^2+\beta_2^2+\beta_3^2\right)+\alpha_4 \beta_3^2+\alpha_3^3\right) +\notag\\
       &\hskip 0.6cm \alpha_2 \left(\beta_2^2 \left(2 \alpha_3^2+3 \beta_2^2+2 \beta_3^2\right)+
    3 \beta_1^4+6 \beta_2^2 \beta_1^2\right)-\alpha_1^5+5 \alpha_2 \alpha_1^4+
    \alpha_2^5\Bigr)\\
\label{eq:m8}
M_8 = & \beta_1^2 \Bigl(5 \alpha_1^4 \left(3 \alpha_2^2+\beta_1^2+3 \beta_2^2\right) - 
              20 \alpha_1^3 \left(\alpha_2 \left(\beta_1^2+2 \beta_2^2\right)+\alpha_3 \beta_2^2+\alpha_2^3\right) +  \notag\\
   & \hskip 0.6cm \alpha_1^2 \left(15 \alpha_2^2 \left(2 \beta_1^2+3 \beta_2^2\right) + 
              30\alpha_3 \alpha_2 \beta_2^2+15 \beta_2^2 \left(\alpha_3^2+\beta_2^2+\beta_3^2\right) + 
              15 \alpha_2^4+6 \beta_1^4+20 \beta_1^2 \beta_2^2\right) - \notag \\
   & \hskip 0.6cm 2 \alpha_1 \Bigl(2\alpha_2^3 \left(5 \beta_1^2+6 \beta_2^2\right) + 9 \alpha_3 \alpha_2^2 \beta_2^2 + 
              3\alpha_2 \left(\beta_2^2 \left(2 \alpha_3^2+3 \beta_2^2+2 \beta_3^2\right) + 
              2 \beta_1^4 + 5 \beta_2^2 \beta_1^2\right) + \notag\\
              &\hskip 1.4cm \beta_2^2 \left(\alpha_3 \left(5 \beta_1^2+6
              \left(\beta_2^2+\beta_3^2\right)\right)+3 \alpha_4 \beta_3^2+3 \alpha_3^3\right)+3
              \alpha_2^5\Bigr) + \notag\\
   & \hskip 0.6cm 3 \alpha _3^2 \beta _2^4+\alpha _3^4 \beta _2^2+2 \alpha _3^2 \beta _1^2
            \beta _2^2+4 \alpha _2^3 \alpha _3 \beta _2^2+3 \alpha _3^2 \beta _2^2 \beta _3^2+\alpha_4^2 \beta_2^2 \beta_3^2 +  
             2 \alpha _3 \alpha _4 \beta _2^2 \beta _3^2 + 5 \alpha _2^4\left(\beta _1^2+\beta _2^2\right) + \notag\\
    &\hskip 0.6cm 3 \alpha _2^2 \left(\beta _2^2 \left(\alpha _3^2+2 \beta_2^2 + \beta _3^2\right) + 
            2 \beta_1^4 + 4 \beta_2^2 \beta_1^2\right) + 
           2 \alpha_2 \beta_2^2\left(\alpha_3 \left(3 \beta_1^2+3 \beta_2^2+2 \beta_3^2\right)+\alpha_4 \beta_3^2 +
               \alpha_3^3\right) + \notag\\
    &\hskip 0.6cm  \alpha _1^6-6 \alpha _2 \alpha _1^5+\alpha _2^6+\beta _1^6+\beta
   _2^6+3 \beta _1^2 \beta _2^4+\beta _2^2 \beta _3^4+3 \beta _1^4 \beta _2^2+2 \beta _2^4
   \beta _3^2+2 \beta _1^2 \beta _2^2 \beta _3^2+\beta _2^2 \beta _3^2 \beta _4^2\Bigr),
\end{align}
\end{widetext}
In addition, the scaled moments $R_k=M_k/\sigma^k$ (with $\sigma^2 = M_2 \approx \beta_1^2$) can easily be computed using these formulae with the substitutions $\alpha_i \rightarrow \alpha_i/|\beta_1|$ and  $\beta_i \rightarrow \beta_i/|\beta_1|$.

The validity of Eqs. (\ref{eq:h1})-(\ref{eq:m8}) is shown in Table \ref{tab:moments}, where the moments extracted from the first four Lanczos (L) iterations from a single random pivot are compared with the exact (Ex) moments for several nuclei within the $1p0f$-shell model space using the GXPF1A interaction~\cite{gxpf1a}. These systems were chosen because they have large dimensions, $N\approx 2-4\times 10^4$, but are still small enough to fully diagonalize.  For $M_{3-8}$, we show the scaled moments $R_k = M_k/\sigma^k$. Overall, good agreement is obtained between the exact and Lanczos-inferred moments. Some differences exist, which tend to be larger for the higher moments, and are due to an imperfect cancellation in the remainder term that propagates further into the higher moments.  We find, however, that the remainders in $H_1$ and $M_2$ decrease with increasing model space size. We find that these inferred moments are more than sufficient to describe the averaged properties of the Hamiltonian matrix and to model the average properties of the remaining Lanczos matrix elements.

In general, most systems within the $1p0f$ shell have been found to have $R_4 \approx 2.8$, $R_6 \approx 12$, and $R_8 \approx 65-75$. For the purpose of comparison, note that for a Gaussian distribution, $R_4 = 3$, $R_6 = 15$, and $R_8 = 105$. 

\begin{table}
\caption{Comparison between exact (Ex) moments and those computed with the first four Lanczos (L) iterations for selected nuclei in the $1p0f$-shell model space using the GXPF1A interaction. $H_1$ is in units of MeV, $M_2$ is units of MeV$^2$, while $R_{3-8}$ are dimensionless.}
\begin{ruledtabular}
\begin{tabular}{llrrrrrr}
    &   &  $^{47}$Cr  &  $^{47}$Cr  &  $^{48}$Cr   &   $^{48}$Cr   &   $^{72}$Kr  &   $^{73}$Kr  \\
    &   &  $1/2^-$  &  $3/2^-$   &  0$^+$    &  12$^+$   &  0$^+$  &  $1/2^-$   \\
\hline
$H_1$ & Ex  &  -46.326    &    -46.402  &  -55.004 & -59.195  &  -363.738  &  -380.331 \\
          & L   &   -46.335    &    -46.401 &  -54.996 & -59.166  &  -363.695  &  -380.364 \\
\hline
$M_2$ & Ex &  94.722   &  94.052 &  111.121 & 76.011 &  110.502 & 95.473   \\
            & L  &  94.766   &  93.284 &  111.828 & 75.645 &  110.853 & 97.063   \\
\hline
$R_3$  & Ex &  -0.067   &  -0.070 &  -0.072 & -0.092 &  0.021 & 0.039   \\
                      & L    &  -0.089   &  -0.066 &  -0.067 & -0.100 &  0.026 & 0.071   \\
\hline
$R_4$  & Ex &  2.756   &  2.753  &  2.803  &  2.737  &  2.768  &  2.723  \\
                      & L   &  2.763   &  2.780  &  2.777  &  2.765  &  2.763  &  2.710   \\
\hline
$R_5$  & Ex &  -0.612  &  -0.644  &  -0.685  & -0.784  &  0.223  & 0.375   \\
                      & L   &  -0.711  &  -0.620  &  -0.703  & -0.817  &  0.234  & 0.535   \\
\hline
$R_6$  & Ex &  11.742  &  11.724  &  12.421  &  11.515  &  11.875  &  11.190  \\
                      & L   &  11.700  &  11.866  &  12.387  &  11.894  &  11.817  &  11.331   \\
\hline
$R_7$  & Ex &  -5.359  &  -5.706  &  -6.505  & -6.533  &  2.217  &  3.325   \\
                      & L   &  -5.436  &  -5.457  &  -7.776  & -6.656  &  1.916  &  3.930   \\
\hline
$R_8$  & Ex &  65.370   &  65.441  &  74.272  &  63.201  &  66.940  &  59.537  \\
                      & L   &  63.491   &  65.525  &  77.997  &  67.283  &  66.255  &  61.830   \\
\end{tabular}
\end{ruledtabular}
\label{tab:moments}
\end{table}

\begin{figure}
\includegraphics[scale=0.37]{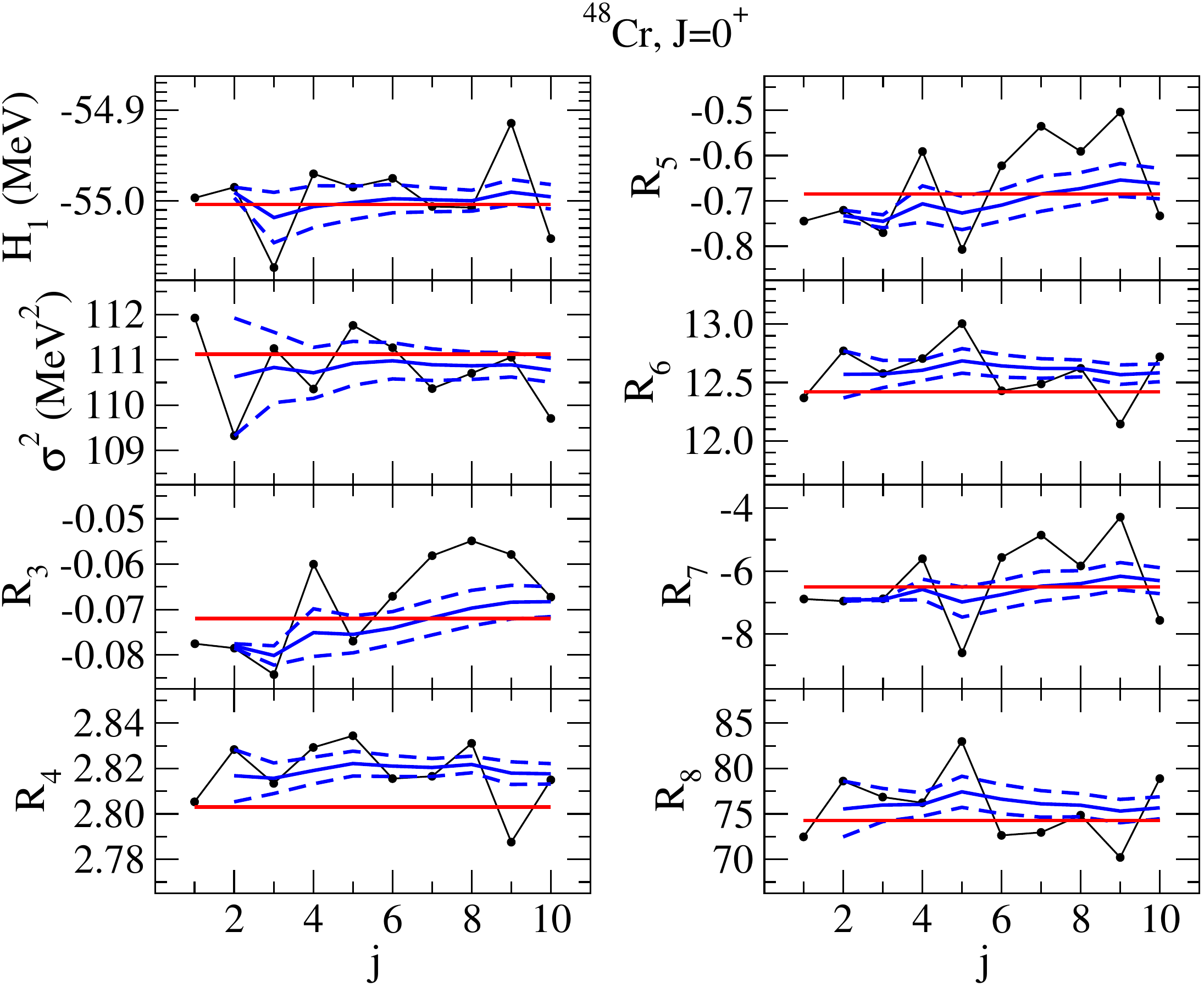}
\caption{(color online) Moments ($H_1$, $\sigma$, and $R_{3-8}$) computed with 10 initial random pivots for the $J^\pi = 0^+$ basis in $^{48}$Cr. The results for each initial vector $v_1^j$ are indicated with the black dots connected with the black line and labeled on the $x$-axis by $j$. The solid blue line represents the running average for each moment, the dashed blue shows the error in the averaging, and the solid red line is the exact result.  }
\label{fig:Cr48_J0}
\end{figure}

As mentioned above, higher accuracy can be achieved by computing the moments stochastically; that is by using $N_{\rm samp}$ different initial pivots $| v_1^j \rangle$ and averaging the resulting moments, i.e.,
\begin{equation}
M_k \approx \frac{1}{N_{\rm samp}} \sum_j \langle v_1^j | \hat h^k | v_1^j \rangle.
\end{equation}
The variance divided by the square root of the number of samples then provides an estimate the error. This is shown in Figure~\ref{fig:Cr48_J0} for the $J^\pi = 0^+$ basis in $^{48}$Cr for $N_{\rm samp} = 10$ different initial random pivots (each sample is indicated by the black dots connected with the black line and labeled on the $x$-axis by the index $j$) for $H_1$, $\sigma$, and $R_{3-8}$ (labeled as $R_{3-8}$ in the figure). The solid blue line represents the running average for each moment, the dashed blue line shows the error in the averaging, and the solid red line is the exact result. The figure shows that for this relatively small system, any single initial pivot provides result with an accuracy of a few percent.

\begin{figure}
\includegraphics[scale=0.37]{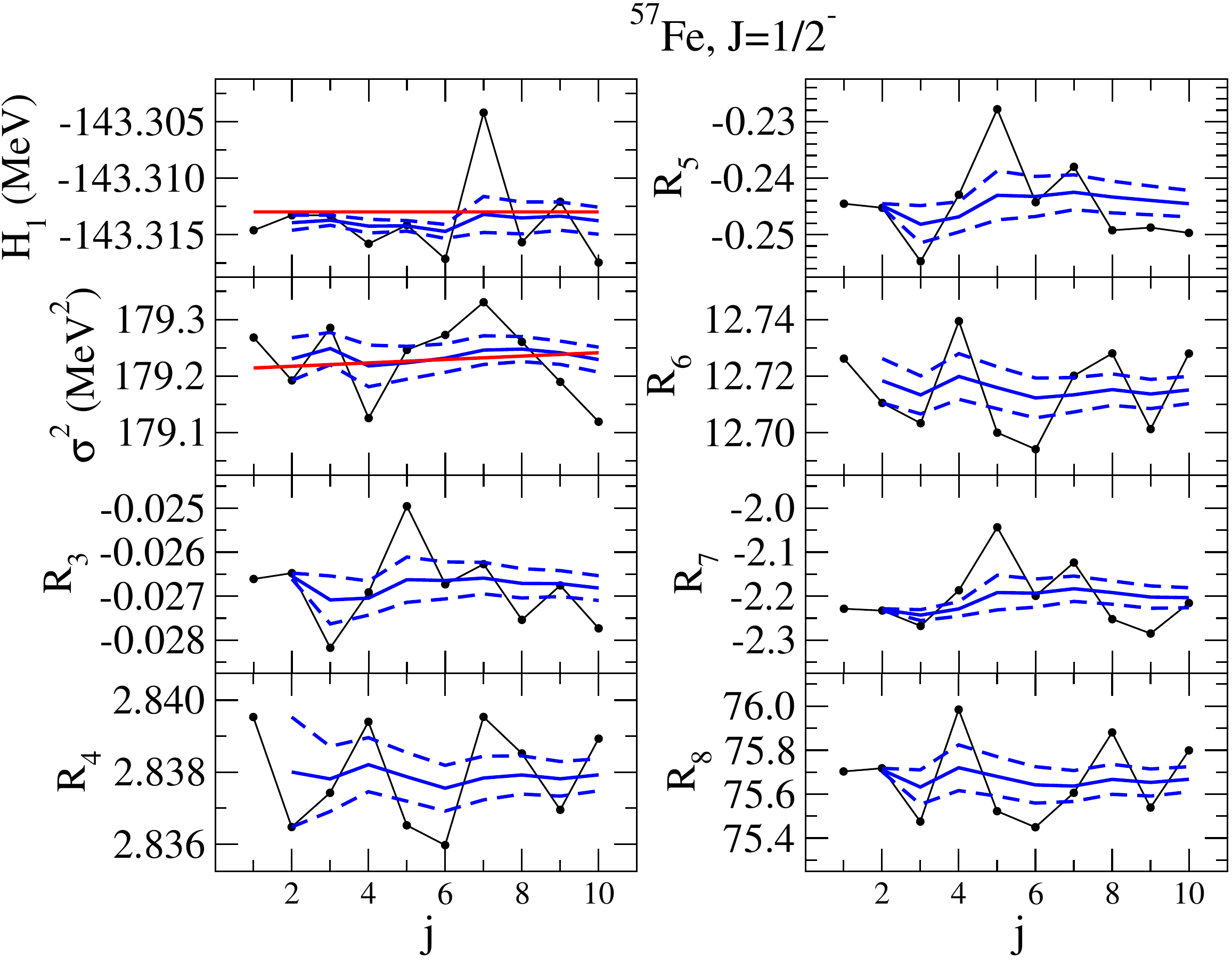}
\caption{(color online) Moments ($H_1$, $\sigma$, and $R_{3-8}$) computed with 10 initial random pivots for the $J^\pi = 1/2^-$ basis in $^{57}$Fe.  The results for each initial vector $v_1^j$ are indicated with the black dots connected with the black line and labeled on the $x$-axis by $j$. The solid blue line represents the running average for each moment, the dashed blue shows the error in the averaging, and the solid red line is the exact result for $H_1$ and $\sigma^2$.  }
\label{fig:Fe57_J1}
\end{figure}

In Figure~\ref{fig:Fe57_J1}, we show moments extracted for 10 different initial random pivots for the $J^\pi = 1/2^-$ states in $^{57}$Fe. Again, the individual results are represented by the black points, while the solid and dashed blue lines represent the running average and the estimated error, respectively. We note that because of the large dimension of this system, $N = 13436903$, the variation in the individual samples is quite small; amounting to less than one percent. The exact results for $H_1$ and $\sigma^2$, as computed with the computer code of Ref.~\cite{Horoi}, are shown with the red lines. Each of the initial pivots agree with $H_1$ to within 10 keV and $\sigma$ to within 5 keV, and the averaged moments are in excellent agreement with the exact result. This demonstrates that the Lanczos procedure to compute the moments improves with dimension.

\begin{figure}
\includegraphics[scale=0.33]{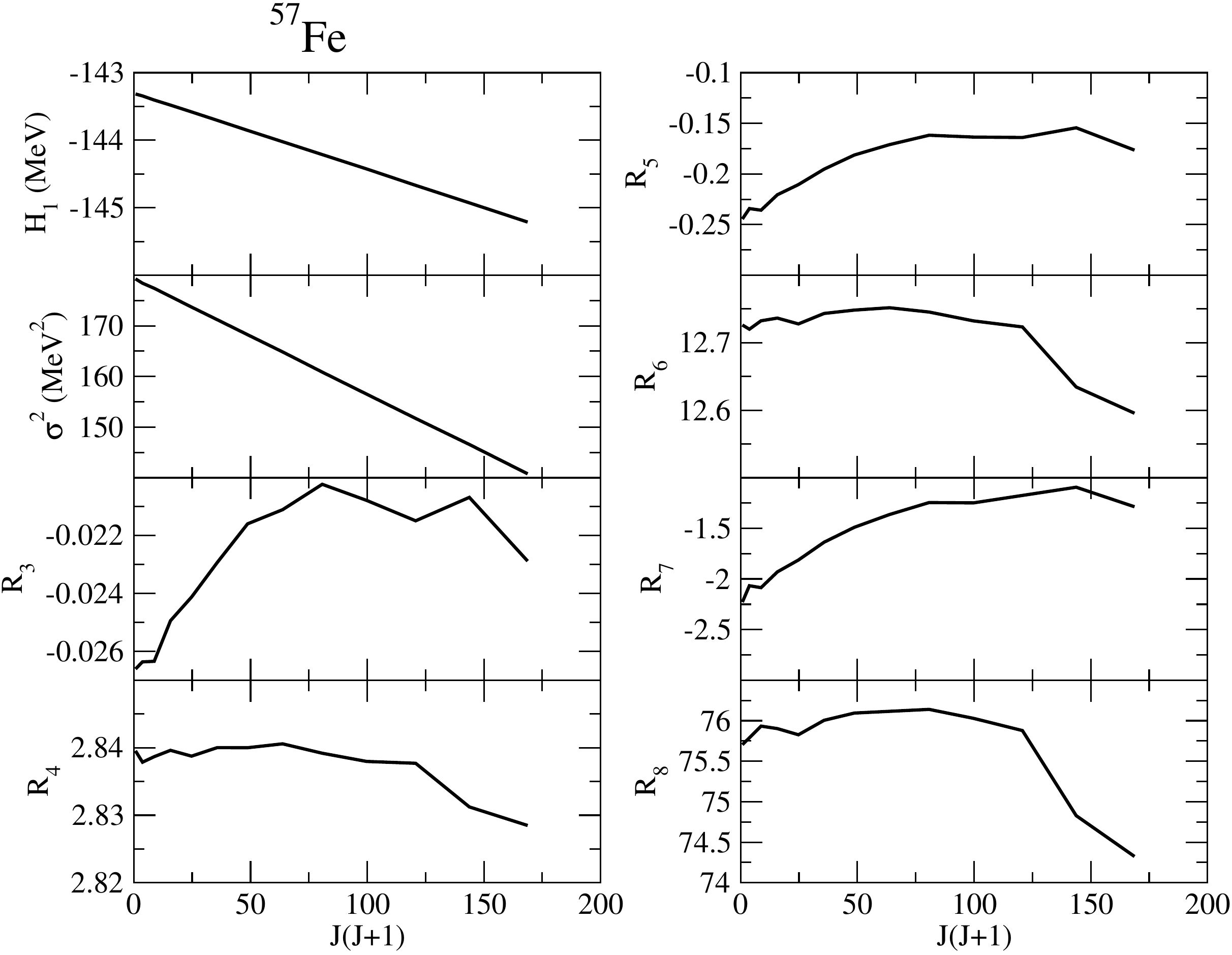}
\caption{(color online) The $^{57}$Fe moments ($H_1$, $\sigma$, and $R_{3-8}$) as a function of the square of the angular momentum $J(J+1)$.  }
\label{fig:Fe57_M_J}
\end{figure}

\begin{figure}
\includegraphics[scale=0.33]{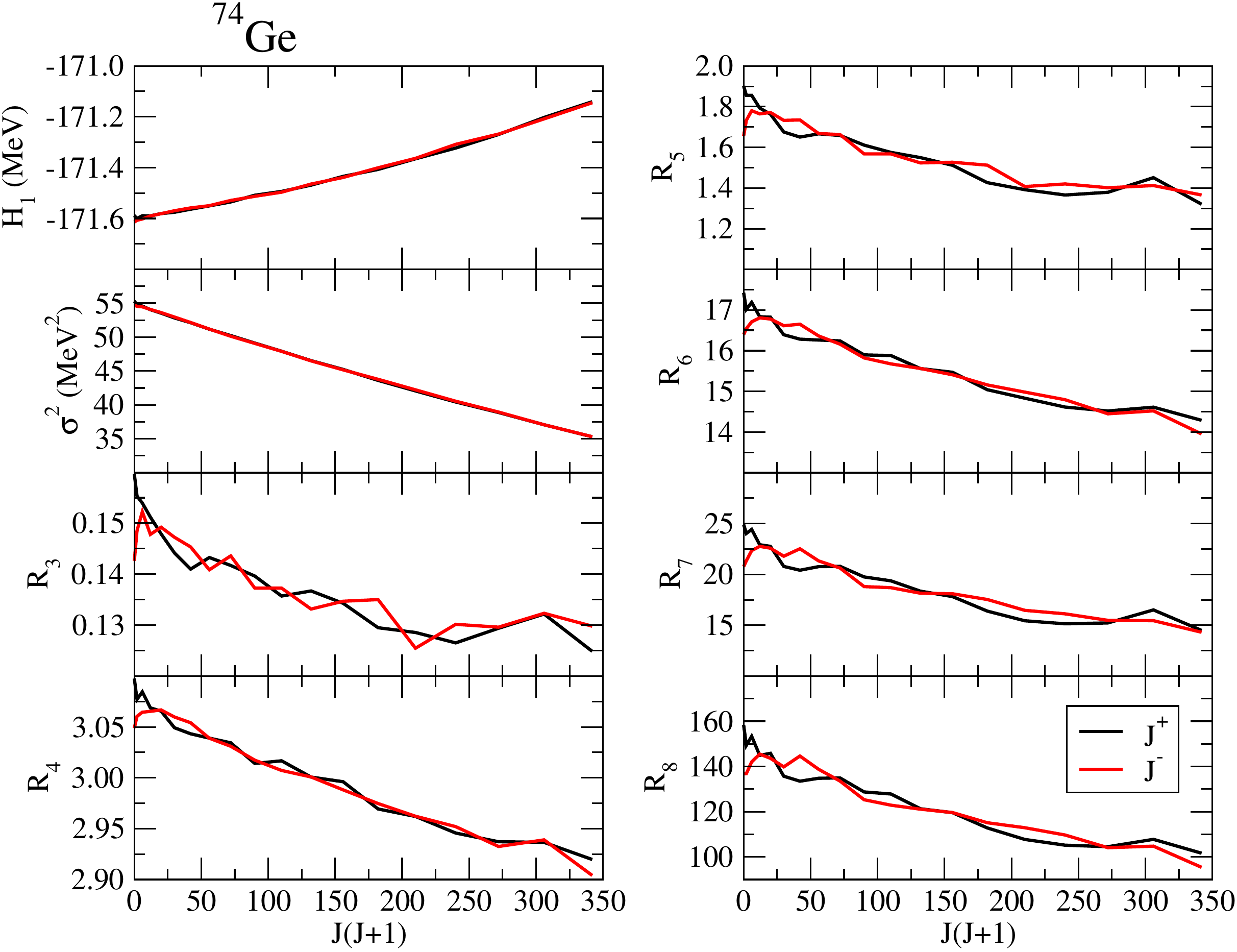}
\caption{(color online) The $^{74}$Ge moments ($H_1$, $\sigma$, and $R_{3-8}$) as a function of the square of the angular momentum $J(J+1)$. The black line shows the dependence for positive-parity states ($J^+$), while the red line shows the negative-parity states ($J^-$). }
\label{fig:Ge74_M_J}
\end{figure}

In Figures~\ref{fig:Fe57_M_J} and \ref{fig:Ge74_M_J}, the dependence on angular momentum [in particular, the square $J(J+1)$] of the first eight moments is shown for calculations of both $^{57}$Fe and $^{74}$Ge. The $^{74}$Ge results were obtained with the  $jj44b$ interaction of Ref.~\cite{Muk}. A strong dependence on the square of the angular momentum is demonstrated for both the first and second moments for both nuclei. For $^{57}$Fe, the scaled higher moments $R_k$ exhibit a weak additional dependence on angular momentum. On the other hand, in $^{74}$Ge, the higher scaled moments show a marked decrease with increasing angular momentum. Indeed, the eigenspectrum transitions to a more Gaussian-like distribution since $R_8$ decreases from a large value of 150 to 100. Also, we note that for low angular momenta $R_4 > 3$. Lastly, the moments for the positive- and negative-parity spaces are nearly identical.

\section{Modeling the Lanczos Matrix Elements}
\label{sec:Modeling}
For large dimensions (e.g., $>10^8$), the computation effort for a shell-model calculation is determined by the Lanczos method; in particular the application of the Hamiltonian to the pivot vectors to generate the tri-diagonal matrix. The resulting tri-diagonal matrix with dimensions of $10^{1-3}$ can easily be diagonalized in a few seconds, while a tri-diagonal matrix with a dimension of the order $10^5$ can be diagonalized within a few minutes. Thus, our goal is to  develop a  method to model the entire tri-digaonal matrix based on the first eight moments. We propose the polynomial form defining the Lanczos matrix elements at each iteration $i$ as
\begin{align}
\label{eq:model_alpha}
\alpha _i =& a_0 + a_1 z_i + {a_2} z_i^2 + {a_3} z_i^3 \\ 
\label{eq:model_beta}
\beta ^2_i=& b_1 z_i[1+ {b_2} z_i + {b_3}z_i^2 + {b_4}z_i^3],
\end{align}
where $  z_{i} = {\rm ln}(i/N)$. We note that this representation is different from the inverse binomial of Ref.~\cite{zuker} and the shifted Gaussian of Ref.~\cite{ormand}. This representation provides the flexibility to accurately model the Lanczos matrix elements  for a wide range of systems including those where the scaled fourth moment is greater than the Gaussian limit, $R_4 > 3$, as is encountered with Ge isotopes. In addition, the large $N$ limit leads to useful analytic formulae for the moments that can be useful to fix the parameters. 

The $a$- and $b$-coefficients can determined by requiring that the moments of the modeled matrix elements reproduce moments of the Hamiltonian. We note that while the moments are in general high-order polynomials in the $a$-  and $b$-parameters, they are, themselves, most sensitive to the odd and even moments, respectively. Further, the dominant parameter is $b_1$, which effectively determines the second moment $M_2$. Also, $a_0$ is trivially constrained by $H_1$ since it does not affect any of the higher moments.  Lastly, we note that many systems (although not all as, is observed later for $^{76}$Ge) have nearly the same value for $b_2$. This is due to the fact, as seen in Table~\ref{tab:moments}, that $R_4\approx 2.7-2.8$, which is close the Gaussian limit of 3.

The first eight moments of the tri-diagonal matrix can be computed via
\begin{align}
\label{eq:hh1}
H_1 = & \langle \alpha \rangle \\
\label{eq:mm2}
M_2 = & \langle (\alpha - \langle \alpha \rangle )^2\rangle + 2\langle \beta^2 \rangle \\
\label{eq:mm3}
M_3 \approx & \langle (\alpha - \langle \alpha \rangle)^3 \rangle + 6\langle (\alpha - \langle \alpha \rangle)\beta^2\rangle \\
\label{eq:mm4}
M_4 \approx& \langle (\alpha - \langle \alpha \rangle)^4 \rangle + 12\langle (\alpha - \langle \alpha \rangle)^2\beta^2\rangle +
       6\langle \beta^4 \rangle\\
\label{eq:mm5}
M_5 \approx& \langle (\alpha - \langle \alpha \rangle)^5 \rangle + 20\langle (\alpha - \langle \alpha \rangle)^3\beta^2\rangle + \notag \\
       & 30\langle (\alpha - \langle \alpha \rangle)\beta^4\rangle\\
\label{eq:mm6}
M_6 \approx& \langle (\alpha - \langle \alpha \rangle)^6 \rangle + 30\langle (\alpha - \langle \alpha \rangle)^4\beta^2\rangle + \notag \\
       & 90\langle (\alpha - \langle \alpha \rangle)^2\beta^4\rangle + 20\langle \beta^6 \rangle\\
\label{eq:mm7}
M_7 \approx& \langle (\alpha - \langle \alpha \rangle)^7 \rangle + 42\langle (\alpha - \langle \alpha \rangle)^5\beta^2\rangle + \notag \\
       & 210\langle (\alpha - \langle \alpha \rangle)^3\beta^4\rangle + 140\langle (\alpha - \langle \alpha \rangle)\beta^6\rangle
\end{align}
\begin{align}
\label{eq:mm8}
M_8 \approx& \langle (\alpha - \langle \alpha \rangle)^8 \rangle + 56\langle (\alpha - \langle \alpha \rangle)^6\beta^2\rangle + \notag \\
       & 420\langle (\alpha - \langle \alpha \rangle)^4\beta^4\rangle + 560\langle (\alpha - \langle \alpha \rangle)^2\beta^6\rangle + \notag \\
       & 70\langle \beta^8 \rangle,   
\end{align}
where $\langle ...\rangle \rightarrow \frac{1}{N} \sum_i ...$, which for large $N$ can be extended to the integral  $\frac{1}{N} \int_1^N ... dx$. The approximate equality arises from the assumption that adjacent matrix elements $\beta_i$, $\beta_{i\pm1}$, $\beta_{i\pm2}$, $\beta_{i\pm3}$ are nearly equal. With Eqs.~(\ref{eq:hh1})-(\ref{eq:mm8}) the $a$- and $b$-parameters can be ``fit'' to reproduce the moments of the Hamiltonian; leading to a modeled tri-diagonal matrix with the same moments as the original Hamiltonian.

In principle, analytic formulae can be obtained for the moments in the large $N$ limit since 
\begin{equation}
\lim_{N \rightarrow \infty} \int_1^N \ln^mxdx = m!.
\end{equation}
In this limit, the first five moments as defined in Eqs.~(\ref{eq:hh1})-(\ref{eq:mm5}) are given in terms of the $a$- and $b$-parameters of Eqs.~(\ref{eq:model_alpha}) and (\ref{eq:model_beta}) by
\begin{widetext}
\begin{align}
\label{eq:ah1}
H_1 =& a_0 - a_1 + 2 a_2 - 6 a_3 , \\
\label{eq:am2}
M_2 =& a_1^2+\left(36 a_3-8 a_2\right) a_1+4 \left(5 a_2^2-54 a_3 a_2+171 a_3^2\right) + 
             2 b_1 \left(2b_2-6 b_3+24 b_4-1\right) \\
\label{eq:am3}
M_3 = &   -2 \Bigl[
                           a_1^3-18 \left(a_2-6a_3\right) a_1^2+12 \left(10 a_2^2-135 a_3 a_2+513 a_3^2\right) a_1 - \notag\\
            & \hskip 0.8 cm             8 \left(37 a_2^3-837a_3 a_2^2+7047 a_3^2 a_2-21897 a_3^3\right)
                     \Bigr] + \notag \\
            & 6 b_1 \Bigl[
                                18 a_3 \left(-6 b_2+38 b_3-272 b_4+1\right)+a_1 \left(-4 b_2+18 b_3-96
                                  b_4+1\right)+4 a_2 \left(5 b_2-27 b_3+168 b_4-1\right)
                          \Bigr] \\
\label{eq:am4}
M_4 = 
& 3 \Bigl[
              3a_1^4+\left(552 a_3-80 a_2\right) a_1^3 + 
              8 \left(113 a_2^2-1746 a_3 a_2+7515 a_3^2\right)a_1^2 - \notag\\
&\hskip 0.4cm              32 \left(158 a_2^3-4059 a_3 a_2^2+38412 a_3^2 a_2-132921 a_3^3\right) a_1 + \notag\\
&\hskip 0.4cm              16\left(731 a_2^4-27540 a_3 a_2^3+427014 a_3^2 a_2^2-3208572 a_3^3 a_2+9800919 a_3^4\right) 
             \Bigr]+ \notag\\
&            12 b_1 \Bigl[
                                  a_1^2 \left(14 b_2-78 b_3+504 b_4-3\right) - 
                                  4 a_1 \Bigl(a_2 \left(44b_2-282 b_3+2064 b_4-8\right) - 9 a_3 \left(32 b_2-234 b_3+1928 b_4-5\right)\Bigr) + \notag\\
&\hskip .9cm                                4\Bigl(a_2^2 \left(158 b_2-1146 b_3+9384 b_4-25\right) - 
                                         36 a_3 a_2 \left(65 b_2-531 b_3+4844b_4-9\right) + \notag\\
&\hskip 0.9cm                                         9 a_3^2 \left(1082 b_2-9846 b_3+99144 b_4-133\right)
                                    \Bigr)
                          \Bigr] + \notag\\
&              12b_1^2 \Big[12 b_2^2-6 \left(20 b_3-120 b_4+1\right) b_2+360 b_3^2 + 20160 b_4^2 + 
                                   b_3\left(24-5040 b_4\right)-120 b_4+1\Bigr] 
\end{align}
\begin{align}
\label{eq:am5}
M_5 =
&  -4 \Big[
     11 a_1^5+10 \left(43 a_2-342 a_3\right) a_1^4 - 
      20 \left(371a_2^2-6507 a_3 a_2+31410 a_3^2\right) a_1^3 + \notag\\
&\hskip 0.8cm      40 \left(1756 a_2^3-50625 a_3 a_2^2+532332a_3^2 a_2-2029563 a_3^3\right) a_1^2 - \notag\\
&\hskip 0.8cm      80 \left(4534 a_2^4-189909 a_3 a_2^3+3245859 a_3^2a_2^2-26685153 a_3^3 a_2 + 
                  88602417 a_3^4
           \right)a_1 + \notag\\
&\hskip 0.8cm      32 \left(25411 a_2^5-1442205 a_3a_2^4+35446860 a_3^2 a_2^3 - 
                  469283490 a_3^3 a_2^2+3331562805 a_3^4 a_2-10104948693a_3^5
           \right)
     \Bigr] + \notag\\
&    20 b_1\Bigl[
                         a_1^3 \left(-64 b_2+426 b_3-3216 b_4+11\right) + \notag\\
&\hskip 0.9cm  6 a_1^2 \Bigl(2 a_2 \left(119 b_2-891b_3+7488 b_4-18\right) - 
                                             9 a_3 \left(202 b_2-1694 b_3+15792 b_4-27\right)
                                        \Bigr) - \notag\\
&\hskip 0.9cm  12 a_1\Bigl(a_2^2 \left(988 b_2-8238 b_3+76416 b_4-133\right) - 
                                            36 a_3 a_2 \left(466 b_2 - 4313b_3+44036 b_4-56\right) + \notag\\
&\hskip 1.9cm                      9 a_3^2 \left(8764 b_2-89298 b_3+996336 b_4-949\right)
                                     \Bigr) + \notag\\
&\hskip 0.9cm   8\Bigl(a_2^3 \left(4534 b_2-41754 b_3+424416 b_4-548\right) - 
                                    27 a_3 a_2^2 \left(4714b_2-47818 b_3+531392 b_4-513\right) + \notag\\
&\hskip1.3cm              54 a_3^2 a_2 \left(24245 b_2-268947 b_3+3246768b_4-2395\right) + \notag\\
&\hskip1.3cm              27 a_3^3 \left(-181498 b_2+2187714 b_3-28528896b_4+16391\right)
                          \Bigr)
              \Bigr] - \notag\\
& 120 b_1^2 \Bigl[
                                 -a_2 \left(168 b_2^2-6 \left(400 b_3-3240 b_4+9\right) b_2 + 
                                                 9720b_3^2+887040 b_4^2-2400 b_4 - 
                                                 336 b_3 \left(525 b_4-1\right)+5
                                          \right) + \notag\\
&\hskip 1cm               a_1 \Bigl(24 b_2^2 - 3\left(100 b_3-720 b_4+3\right) b_2 + 
                                                1080 b_3^2 + 80640 b_4^2 - 300 b_4 - 
                                                 24 b_3 \left(735b_4-2\right)+1
                                          \Bigr) + \notag\\
&\hskip 1cm               9 a_3 \Bigl(136 b_2^2 - 2 \left(1100 b_3-9960 b_4+19\right) b_2 + 
                                                    9960b_3^2+1102080 b_4^2-2200 b_4 - \notag\\
&\hskip 1.9cm                                                    272 b_3 \left(735 b_4-1\right) +3 
                                                \Bigr)
            \Bigr]
\end{align}
\end{widetext}
For $k > 5$, these formulae are more complicated with extremely large coefficients. Nonetheless, the analytic formulae for $M_3$ and $M_5$ are useful for providing initial estimates for the parameters $a_1$ and $a_2$. An alternative, that is somewhat more efficient for the higher moments ($k \ge 5$), and was used here to determine the parameters, is to evaluate the moment integrals numerically using $z$ as the integration variable, which involves integrals of the form
\begin{equation}
\frac{1}{N}\int_{\ln(1/N)}^0 e^z z^m dz.
\end{equation}
Sufficient accuracy can be achieved using Simpson's rule with $10^5$ points.  For numerical stability, 
the integrals can be evaluated by scaling relative to $M_2$ by taking $a_i \rightarrow a_i/\sqrt{-b_1}$ followed by setting $b_1 \rightarrow -1$.

The procedure used here to find the $a$- and $b$- parameters is discussed in Appendix~\ref{app:A}.

\begin{figure}
\includegraphics[scale=0.39]{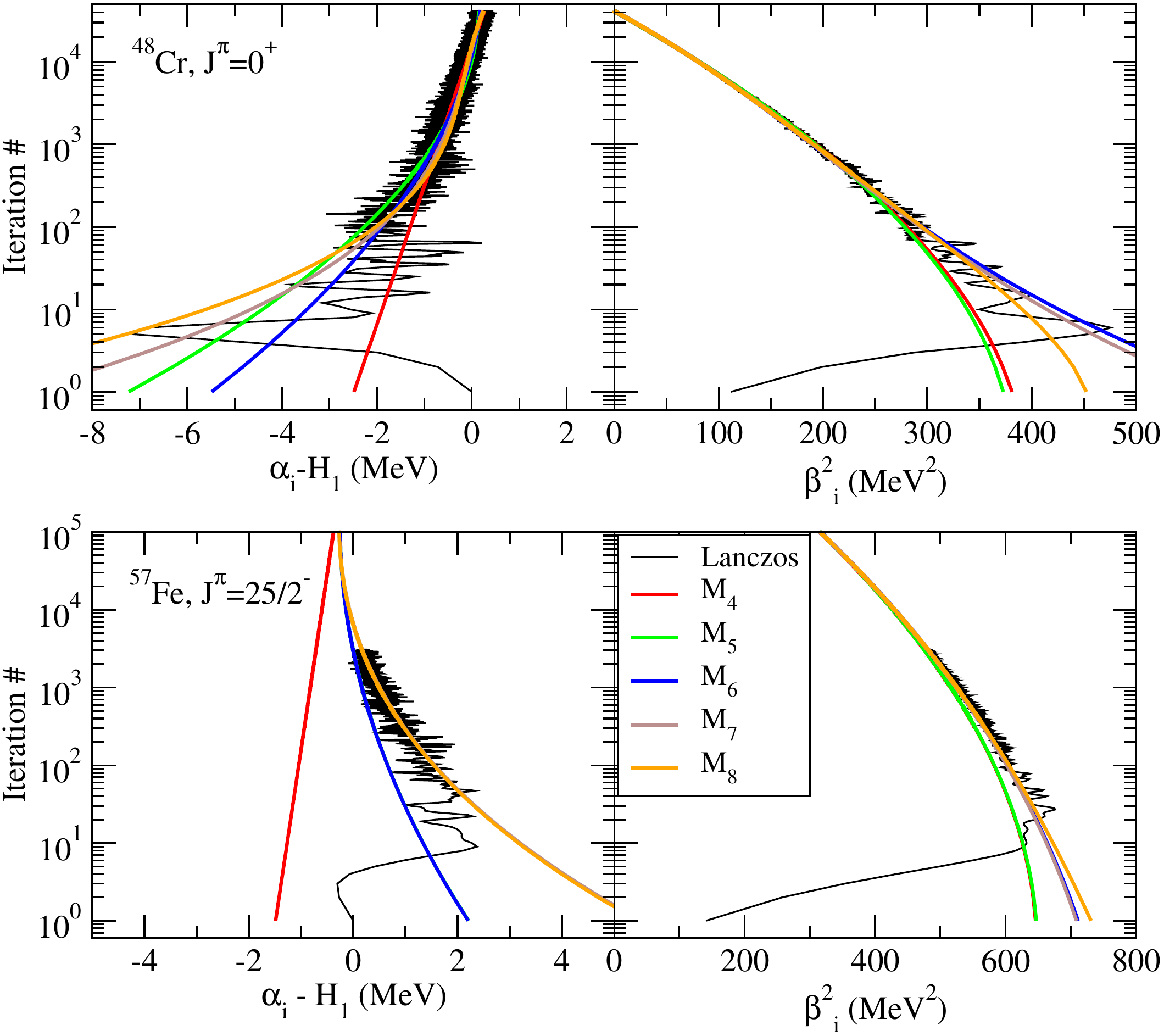}
\caption{(color online) Comparison between shell model (black) and modeled Lanczos matrix elements $\alpha$ and $\beta$ for $^{48}$Cr, $J^\pi = 0^+$ (top) and $^{57}$Fe, $J^\pi=25/2^-$ (bottom) within the $1p0f$-model space using the GXPF1A interaction~\cite{gxpf1a}. The colored curves show modeled Lanczos matrix elements using Eqs.~(\ref{eq:model_alpha}) and (\ref{eq:model_beta}) with the indicated moments to constrain the $a$- and $b$-parameters. }
\label{fig:alpha_beta}
\end{figure}

The utility of the moment method to describe the nuclear Hamiltonian is illustrated in Figure~\ref{fig:alpha_beta} where the modeled (colored lines) Lanczos matrix elements are compared with those obtained from a shell-model calculation (black lines) for the $^{48}$Cr, $J^\pi = 0^+$ (top) and $^{57}$Fe, $J^\pi=25/2^-$ (bottom) systems.  For $^{48}$Cr the entire Lanczos matrix ($N=41355$) is plotted, while for $^{57}$Fe, $J^\pi = 25/2^-$ ($N=13752093$), 3074 Lanczos iterations were performed and 100000 modeled matrix elements are shown. The $^{48}$Cr system is somewhat typical where the dominant behavior observed in the Lanczos matrix can be extracted from just the first four moments, i.e., $M_3$ to constrain $a_1$ and $M_2$ and $M_4$ to constrain $b_2$ and $b_4$. Still, the figure shows that using moments up to $M_8$ can improve the overall description of modeled Lanczos matrix. The $^{57}$Fe system is different in that the higher moments are essential. The figure shows that limiting to $M_3$ to constrain $a_1$ is clearly inadequate and improvement is achieved only by including the higher odd moments, and the best overall results are obtained using all eight moments. The $^{57}$Fe case is also interesting as it has a negative skewness ($M_3$), which is correctly captured with the Lanczos method to compute the moments, but also seemingly contradicts the positive values of ($\alpha_i - H_1$) shown for the first few thousand iterations. Indeed, the diagonal matrix elements show a strong curvature and eventually turn negative for large iteration number. This is captured in the higher odd moments leading to quadratic and cubic terms in the modeled $\alpha_i$ matrix elements. Lastly, the $\beta_i$ at low iteration number are also influenced by the higher even moment $M_6$.

\section{Estimating the Level Density}

The density of states is a key nuclear property that has a significant impact on reaction rates for statistical processes, such as radiative neutron capture. For the most part, reaction models, such as Hauser-Feshbach~\cite{Hauser-Feshbach}, have relied on a parameterization of the level density based on a modified back-shifted Fermi gas approach such as was introduced by Gilbert and Cameron~\cite{Gilbert-Cameron}. This approach requires knowledge about several parameters such as the single-particle level-density parameter $a$, which may depend on excitation energy, the pairing gap $\Delta$, and the spin cutoff parameter. In addition, the back-shifted Fermi gas density is matched to the low-lying spectrum where the level density is assumed to follow an exponential form. The matching is accomplished by requiring that the exponential component reproduces the cumulative density up to an excitation where the discrete levels are both known and complete and requiring continuity in the logarithmic derivate of the level density (equivalent to the inverse temperature) at the matching energy. A drawback of this procedure is that the level-density parameters are generally constrained by experimental knowledge, such as the spacings of $l=0$ ($D_0$) and $l=1$ ($D_1$) resonances at the neutron separation energy, $S_n$. These quantity are generally known only in systems based on a stable target. For radiative neutron capture, the level density is needed essentially up to the neutron separation energy. 

One approach to generalize our knowledge of the level density is to use theoretical structure models based on the microscopic physics involved, such as the nuclear shell model, where high-quality empirical nuclear Hamiltonians have been developed that are well-known to reproduce the low-lying spectra of nuclei. It is important to note that these shell-model calculations are based on a finite model space, and at some excitation energy, $E_x$, they will fail to adequately enumerate the system due to the presence of so-called ``intruder'' states. These intruder states, however, are expected to occur at higher excitation energies, generally of the order of the shell gap for states with opposite parity and twice the shell gap for states of the same parity. Thus, in many cases it is not unreasonable to hope that a large-basis shell-model calculation contains contains sufficient configurations to adequately describe the states of a given parity up to excitation energies near the neutron separation energy. This supposition can be tested in a few cases through comparison with experimentally measured resonance spacings. For example, within the $1p0f$-shell, the calculated density of states can be compared with the $l=1$ spacings $D_1$, at which point, the computed level density can be used to define parameters of the back-shifted Fermi gas needed to describe the full level density. 

The most straight forward approach to compute the density of states within the shell model would be to simply diagonalize the model Hamiltonian and count the respective states. In many cases, this is computationally prohibitive since the number of the configurations within the model space can exceed $10^9$. Instead, since the density of states is more of a statistical property of the Hamiltonian, we propose to model the Hamiltonian via the moments method outlined above and to compute the density of states from the modeled matrix. Another approach would be to use the binomial distribution described in Ref.~\cite{zuker}, which is constrained with just the first four moments of the Hamiltonian and is appealing due to its analytic nature. In what follows, several approaches to determine the density of states as a function of excitation energy are outlined.

\subsection{Extrapolated Lanczos Method}
\label{sec:ELM}
Section \ref{sec:Modeling} illustrated that for most cases the global, or averaged, properties of the Lanczos matrix can be predicted from just four Lanczos iterations. This offers a strategy to predict the statistical properties of the entire energy spectrum by performing a set of Lanczos iterations sufficient to describe the low-lying spectrum and then extrapolate the Lanczos matrix elements with Eqs.~(\ref{eq:model_alpha}) and (\ref{eq:model_beta}) to an iteration number sufficient to properly estimate the density of states. We refer to this as ELM($k$,$N_{\rm Lanc}$), where $k$ denotes the maximum moment $M_k$ used to extrapolate the Lanczos matrix elements and $N_{\rm Lanc}$ is the number of actual Lanczos iterations used prior to extrapolation. In general, the Lanczos iterations can be computationally expensive for large model spaces, and a key question is just what value of $N_{\rm Lanc}$ is sufficient and/or optimal. A general requirement is obtaining sufficient accuracy in the ground-state energy, $E_{gs}$, to establish the excitation energy scale to measure the density of states. The accuracy required in $E_{gs}$ is model space and Hamiltonian dependent. For example, for the model spaces and Hamiltonians studied in this work, we found that an uncertainty of 10 keV in $E_{gs}$ leads to a 1\% uncertainty in the level density, while a 100 keV uncertainty leads to a 10\% change in the level density. As a general rule, 30 - 40 Lanczos iterations are needed to determine the ground-state energy with an accuracy better than 10 keV, and more often than not, with an accuracy of 1 keV. To some degree, an optimal number of Lanczos iterations can be thought of as where a smooth transition (within the fluctuations of the Lanczos matrix elements) occurs between the computed and modeled Lanczos matrix elements. This may not always be practical, and while it is true that too few iterations can lead to difficulties in the direct computation of the level density at lower excitation energies, an analytic continuation method, discussed below, can address this issue. Consequently, it is often possible to achieve excellent results with the ELM method with $N_{\rm Lanc}$ as low as 40.

\begin{figure}
\includegraphics[scale=0.36]{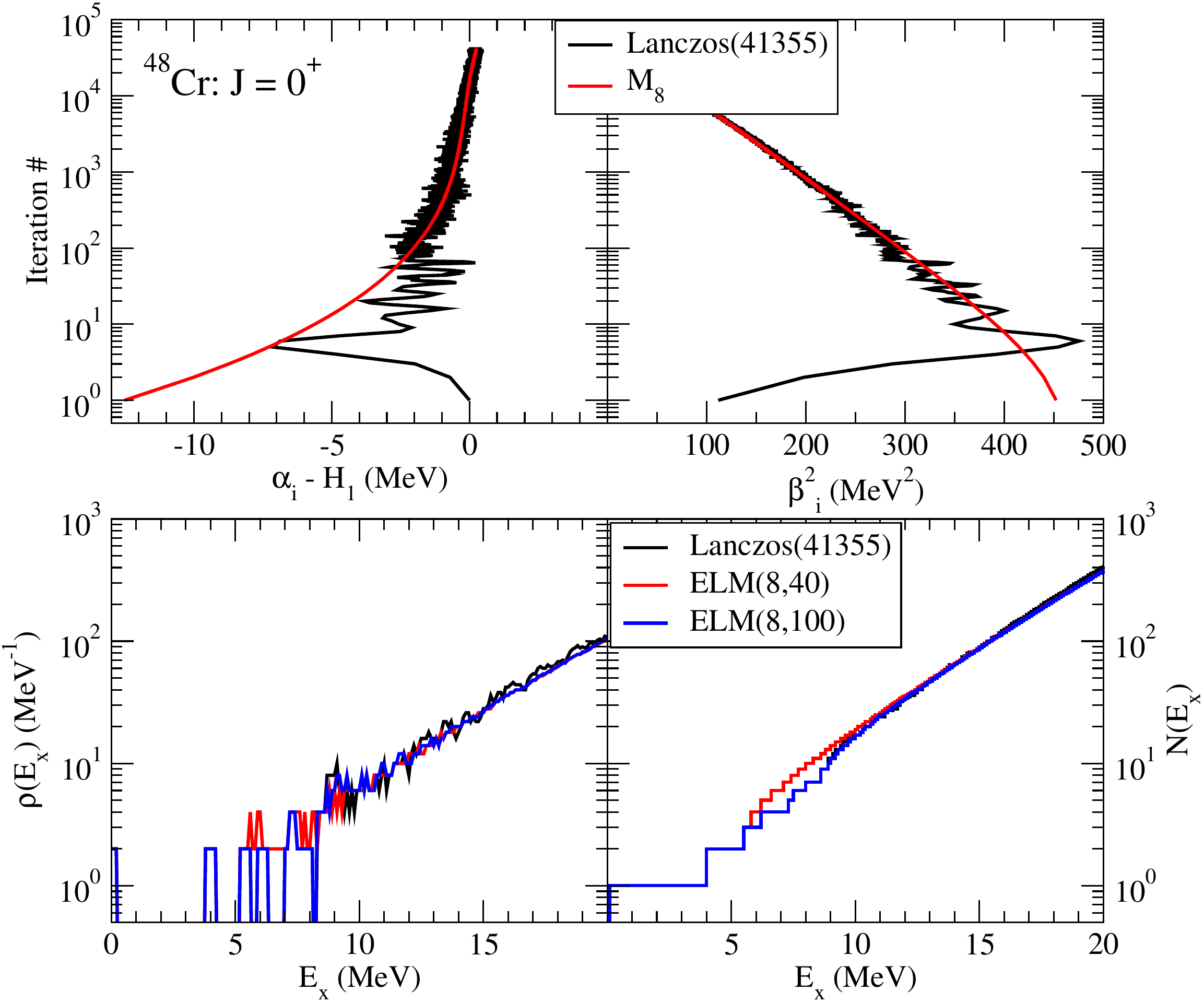}
\caption{
(color online) Results for the $J^\pi=0^+$ space in $^{48}$Cr within the $1p0f$-shell model space with the GXPF1A interaction. The black lines show the shell-model calculations for the Lanczos matrix elements in the upper half of the figure and the level density and cumulative level density in the bottom. In the lower half of the figure the level density and cumulative density are shown for ELM(8,40) (red) and ELM(8,100) (blue).}
\label{fig:ld_fa4800}
\end{figure}

In Figure~\ref{fig:ld_fa4800}, results for the $J^\pi = 0^+$ space in $^{48}$Cr are shown. The shell model calculation was performed using the GXPF1A interaction~\cite{gxpf1a} within the $1p0f$-shell model space with the shell model-code NuShell. Here, the full shell-model matrix was diagonalized with the Lanczos algorithm. The black lines show the results from the shell-model calculation with the Lanczos matrix elements displayed in the top half of the figure and the level density and cumulative density shown in the left and right sides, respectively, in the bottom half of the figure. The level density was computed as function of excitation energy in steps of 100 keV as a running average within an energy window of $E_x \pm 500$ keV, which smooths out fluctuations in the level density. The red and blue lines show the results for ELM(8,40) and ELM(8,100), respectively, where the Lanczos matrix was extrapolated to 50,000 iterations. The ELM(8,100) calculation is nearly indistinguishable from the shell model calculation. The ELM(8,40) calculation shows a slight deviation from the exact shell-model calculation at $E_x\approx 6$~MeV. This deviation is primarily due to a small discontinuity in the matching of the Lanczos matrix elements at $N_{\rm Lanc}$ and hints at how the ELM($k$,$N_{\rm Lanc}$) approach can break down.

In addition to the demonstration for $^{48}$Cr, we have also applied and tested the ELM method to $^{57}$Fe for $J^\pi = 1/2^- - 25/2^-$ and $^{76}$Ge for $J = 0^\pm - 14^\pm$. In what follows, representative results for these systems are shown to demonstrate various features of the ELM method.  We note that applications of the ELM(2,100) method to the Fe region were published earlier in Ref.~\cite{CERN}.

\begin{figure}
\includegraphics[scale=0.36]{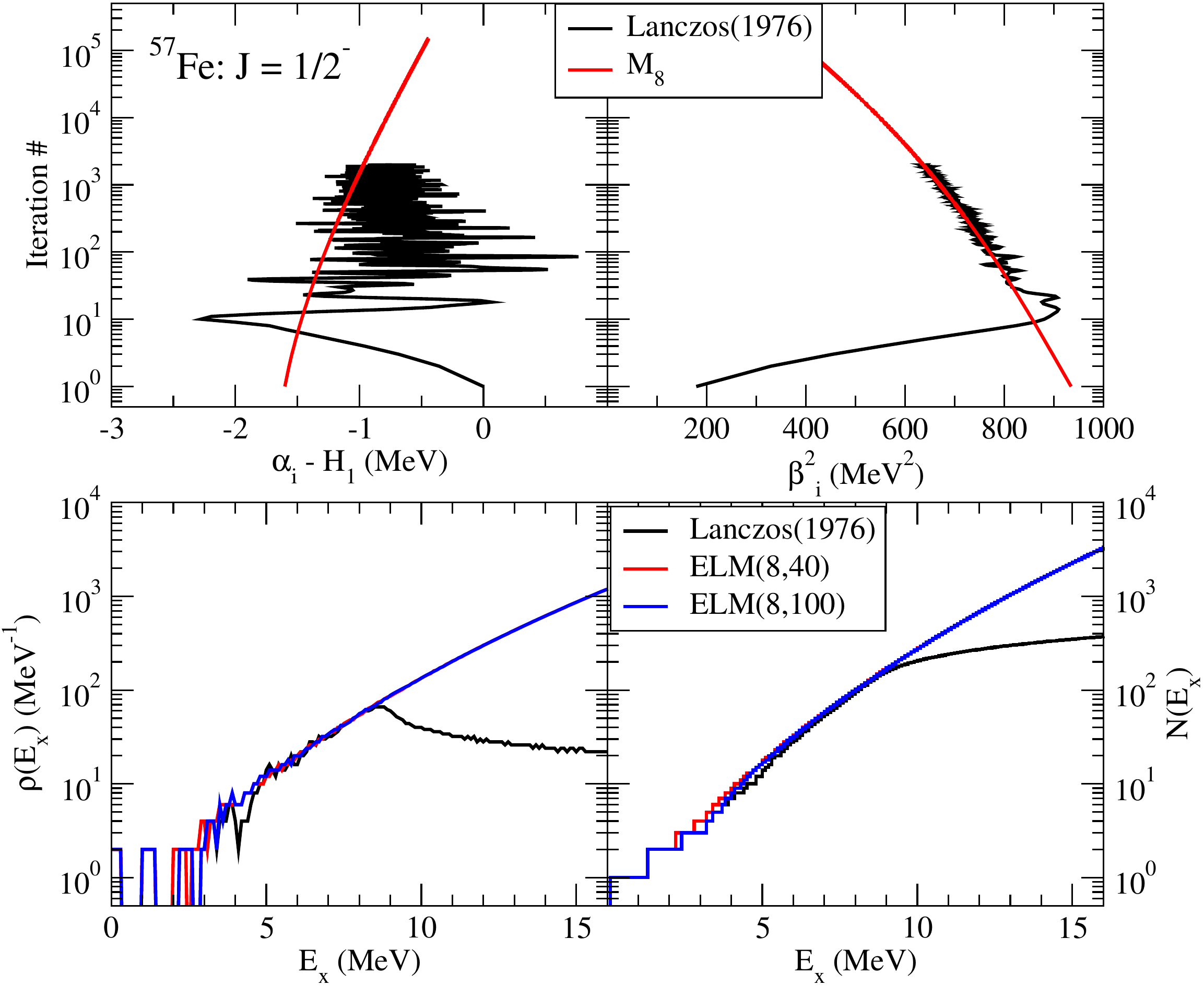}
\caption{
(color online) Results for the $J^\pi=1/2^-$ space in $^{57}$Fe within the $1p0f$ shell-model space with the GXPF1A interaction. The black lines show the shell-model calculations for the Lanczos matrix elements in the upper half of the figure and the level density and cumulative level density in the bottom. In the lower half of the figure the level density and cumulative density are shown for ELM(8,40) (red) and ELM(8,100) (blue). }
\label{fig:ld_fa6h11}
\end{figure}

\begin{figure}
\includegraphics[scale=0.36]{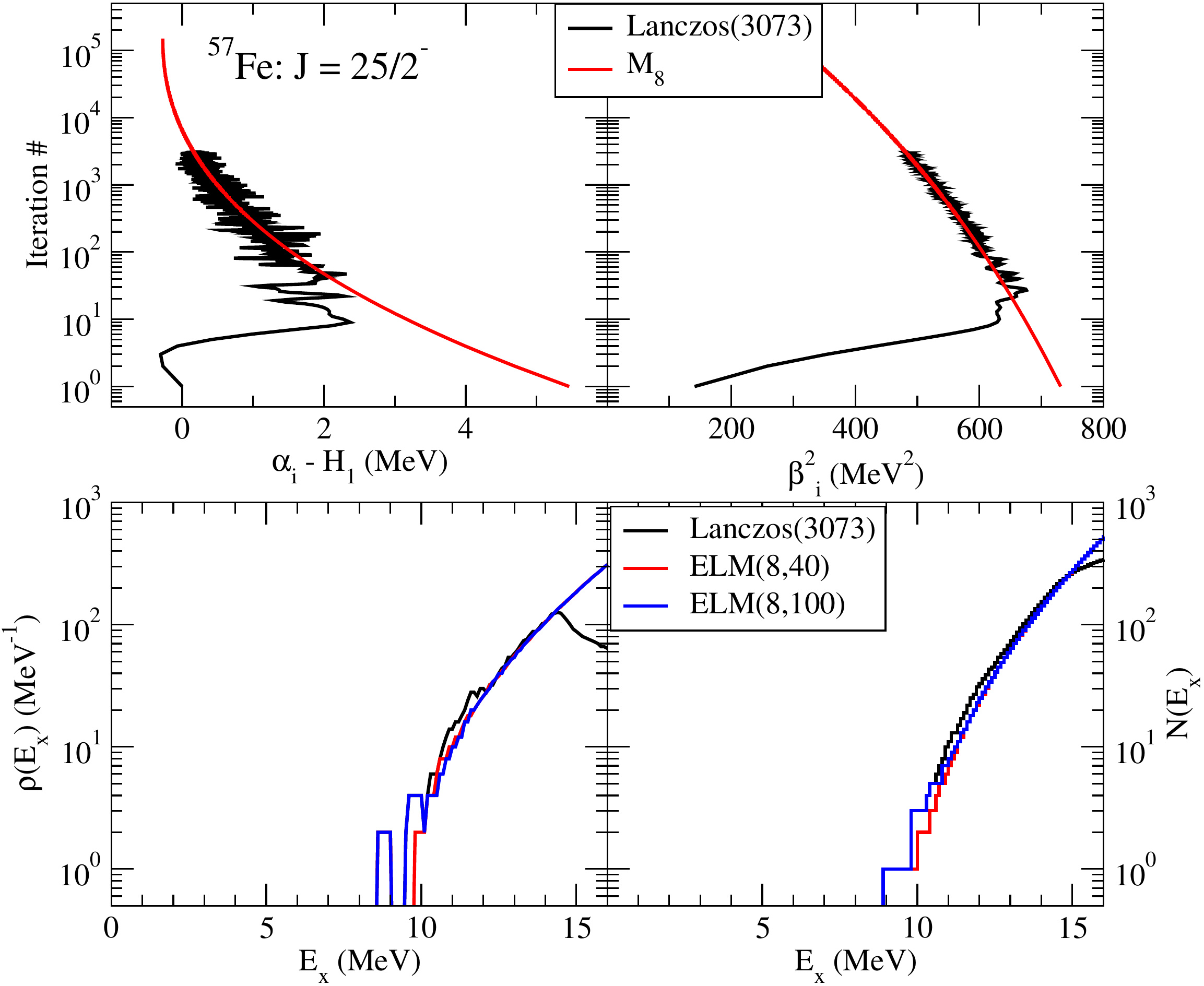}
\caption{
(color online) Results for the $J^\pi=25/2^-$ space in $^{57}$Fe within the $1p0f$ shell-model space with the GXPF1A interaction. The black lines show the shell-model calculations for the Lanczos matrix elements in the upper half of the figure and the level density and cumulative level density in the bottom. In the lower half of the figure the level density and cumulative density are shown for ELM(8,40) (red) and ELM(8,100) (blue). }
\label{fig:ld_fa6h1p}
\end{figure}

Shown in Figures \ref{fig:ld_fa6h11} and \ref{fig:ld_fa6h1p} are the results obtained for the $1/2^-$ and $25/2^-$ states in $^{57}$Fe, while the moments are given in Table~\ref{tab:fe57_moments}. Again, the solid black lines are the results fo the shell-model calculation, while the red and blue lines represent the ELM(8,40) and ELM(8,100) results, respectively. The level densities were computed by extrapolating the the Lanczos matrix elements to 150,000 iterations, diagonalizing the resulting matrix, and as a running average over an excitation energy window of $E_x \pm 500$ keV. The primary difference between the $1/2^-$ and $25/2^-$ angular momentum spaces lies with the odd moments. Both systems have nearly identical negative skewness ($R_3$) as is shown in Table~\ref{tab:fe57_moments}. The high-spin state, however, has a large non-linear term, and the ($\alpha_i - H_1$) are actually positive for smaller iteration number, and then decrease and become negative at large iteration number. A signature of this behavior is also exhibited in the higher odd moments. In particular, when $M_3$ dominates the spectral behavior (linear terms in the $\alpha_i$), one often finds $R_7 \sim 9.0-9.5 R_5$ and $R_5 \sim 9.0-9.5 R_3$. Instead, for the $25/2^-$ space $R_7 \sim 7.3 R_5$ and $R_5 \sim 8 R_3$.

\begin{table}
\caption{Comparison of moments computed with the first four Lanczos iterations for $1/2^-$ and $1/2^-$ angular momentum configuration space in $^{57}$Fe.  $H_1$ is in units of MeV, $M_2$ is units of MeV$^2$, while $R_{3-8}$ are dimensionless.}
\begin{ruledtabular}
\begin{tabular}{lrr}
    &    $^{57}$Fe  &  $^{57}$Fe   \\
    &    $1/2^-$  &  $25/2^-$    \\
\hline
$H_1$ &  -143.314    &    -145.213   \\
$M_2$ &  179.268   &  140.764 \\
$R_3$  &  -0.026   &  -0.022    \\
$R_4$  &  2.839   &  2.828   \\
$R_5$  &  -0.244  &  -0.176   \\
$R_6$  &  12.726  &  12.595   \\
$R_7$  &  -2.229  &  -1.287   \\
$R_8$  &  75.703   &  74.324  \\
\end{tabular}
\end{ruledtabular}
\label{tab:fe57_moments}
\end{table}

This section demonstrating the ELM approach is concluded with an examination of the $J^\pi = 0^+$ and $4^+$ systems in $^{76}$Ge using the $jj44b$ interaction of Ref.~\cite{Muk}. The computed moments are shown in Table~\ref{tab:ge76_moments}. The key features of this system are: 1) the large skewness ($R_3 \sim 0.2$), which is an order of magnitude larger than that observed in $^{57}$Fe, 2) the large fourth moment ($R_4 > 3$, which is substantially larger than the Gaussian value of 3), and 3) the dramatic difference in the $8^{th}$ moment between the two angular momenta.

\begin{table}
\caption{Comparison of moments computed with the first four Lanczos iterations for $0^+$ and $4^+$ states in $^{76}$Ge.  $H_1$ is in units of MeV, $M_2$ is units of MeV$^2$, while $R_{3-8}$ are dimensionless.}
\begin{ruledtabular}
\begin{tabular}{lrr}
    &    $^{76}$Ge  &  $^{76}$Ge   \\
    &    $0^+$  &  $4^+$    \\
\hline
$H_1$ &  -190.500    &    -190.544   \\
$M_2$ &  47.911   &  46.021 \\
$R_3$  &  0.228   &  0.201   \\
$R_4$  &  3.266  &  3.135   \\
$R_5$  &  3.079  &  2.441   \\
$R_6$  &  22.298  &  18.436   \\
$R_7$  &  53.417  &  32.914   \\
$R_8$  &  310.668   &  180.656  \\
\end{tabular}
\end{ruledtabular}
\label{tab:ge76_moments}
\end{table}

\begin{figure}
\includegraphics[scale=0.36]{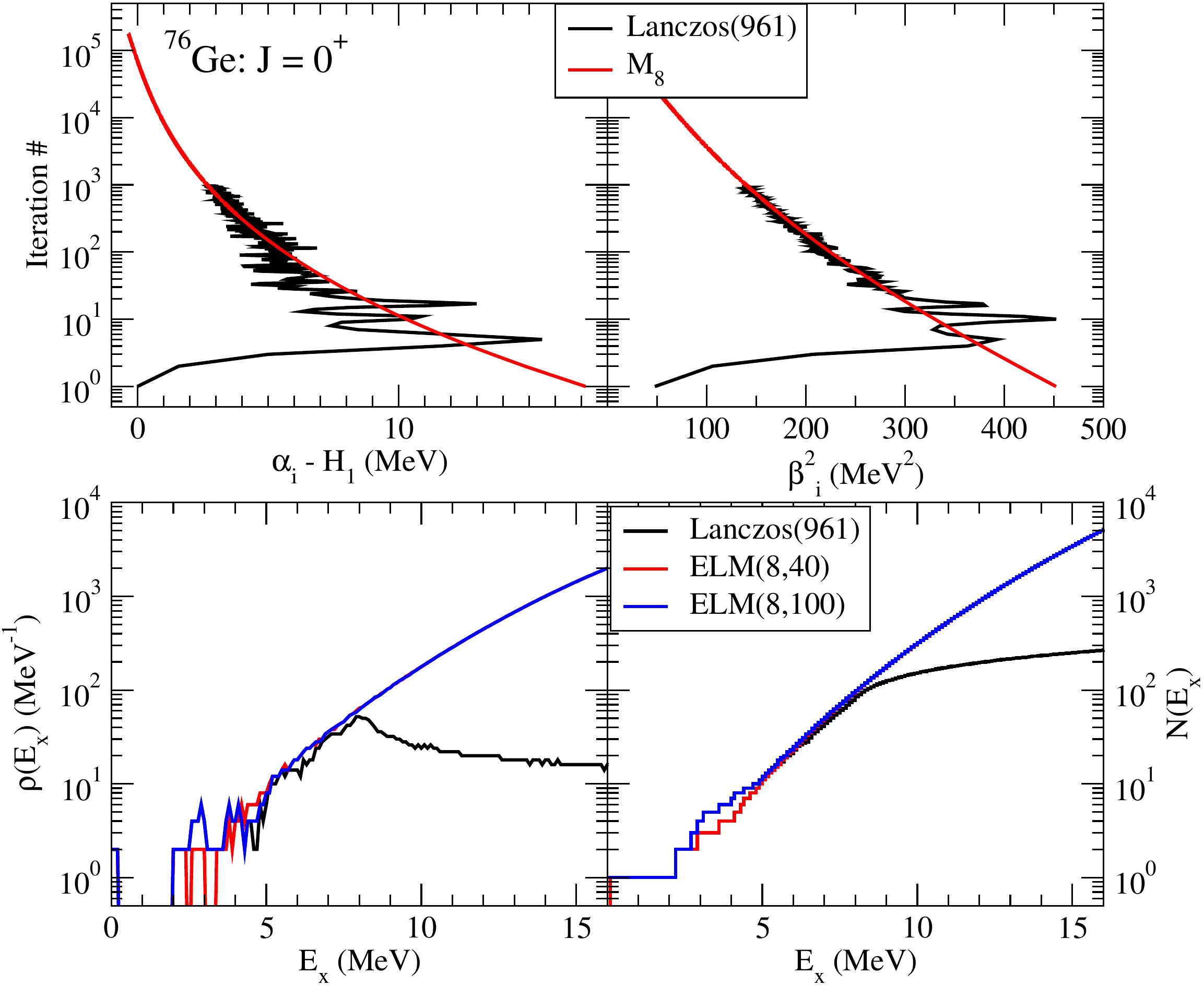}
\caption{
(color online) Results for the $J^\pi=0^+$ space in $^{76}$Ge within the $jj44$ shell-model space with the $jj44b$ interaction. The black lines show the shell-model calculations for the Lanczos matrix elements in the upper half of the figure and the level density and cumulative level density in the bottom. In the lower half of the figure the level density and cumulative density are shown for ELM(8,40) (red) and ELM(8,100) (blue). }
\label{fig:ld_ib4k00}
\end{figure}

\begin{figure}
\includegraphics[scale=0.36]{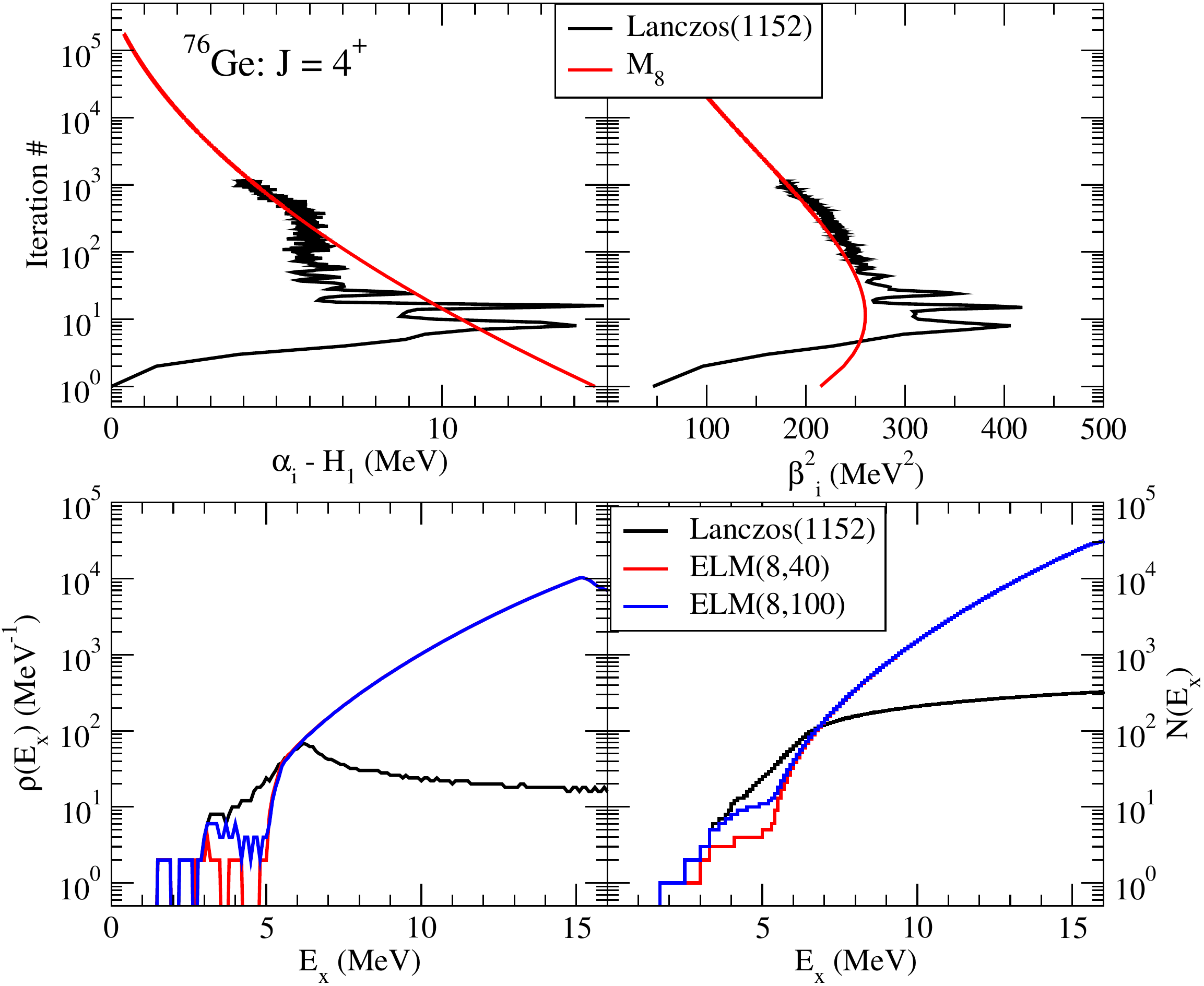}
\caption{
(color online) Results for the $J^\pi=4^+$ space in $^{76}$Ge within the $jj44$ shell-model space with the $jj44b$ interaction. The black lines show the shell-model calculations for the Lanczos matrix elements in the upper half of the figure and the level density and cumulative level density in the bottom. In the lower half of the figure the level density and cumulative density are shown for ELM(8,40) (red) and ELM(8,100) (blue). }
\label{fig:ld_ib4k08}
\end{figure}

Shown in Figures~\ref{fig:ld_ib4k00} and \ref{fig:ld_ib4k08} are the results for $0^+$ and $4^+$ states, respectively, for $^{76}$Ge obtained with the $jj44b$ interaction. The level density was computed by extrapolating the Lanczos matrix to a dimension of 150,000 and computing a running average within the excitation energy of $E_x \pm 500$ keV. For illustrative purposes, approximately 1000 lanczos iterations were performed in each space to to diagnose the calculation in the level density. The results for the $J^\pi=0^+$ space are similar to those shown earlier for $^{48}$Cr and $^{57}$Fe where the ELM(8,100) closely matches the shell-model result. This is not the case, however, for the $J^\pi=4^+$ where there is a clear discrepancy in the spectrum at $E_x \approx 3 - 5$ MeV. On the other hand, the ELM(8,$N_{\rm Lanc}$) results agree with the shell model at higher excitation energies, as would be expected since this is the regime where the statistical nature of the configuration space should dominate the spectral behavior.  The cause of this discrepancy is evident in the upper part of the figure where the diagonal $\alpha_i$ matrix elements exhibit a clear transition in their behavior. The figure shows that the modeled matrix elements capture the overall behavior of the Lanczos matrix elements for large iteration number, but fail to describe the ``step'' behavior shown to at approximately 400 iterations. Thus, the modeled matrix elements lead to a strong dip in the level density for $E_x \approx 3-5$ MeV that is caused by a strong discontinuity between the modeled and actual matrix elements that is far larger than scatter, or noise, exhibited in the computed Lanczos matrix elements. In this case, it would be necessary to perform an ELM(8,400) calculation in order to more accurately describe the system. It has to be noted that often times such a calculation can be computationally prohibitive. In addition, while these calculations for $^{76}$Ge are quite different than those in the $1p0f$ shell, it is not always clear if, or where, a sudden transition in the computed matrix elements may take place; especially for model spaces involving orbits in different major shells. As is apparent from the upper part of Figure~\ref{fig:ld_ib4k08}, the clearest signature of a potential problem with the ELM procedure is the existence of a strong discontinuity at $N_{\rm Lanc}$ between the compute Lanczos matrix elements and the modeled matrix elements. This discontinuity may be present in either the $\alpha_i$ matrix elements, the $\beta_i$ matrix elements, or both. If such a discontinuity exists, two alternatives are suggested: 1) an alternative extrapolation between the computed and modeled matrix elements that smoothly joins the matrix elements to within the ``noise'' in the matrix elements, or 2) a procedure to analytically continue the level density from the high-energy regime to the lowest state in the model space. The latter approach will be discussed in Section~\ref{sec:AC}.

\subsection{Binomial Approximation for the Level Density}

In Ref.~\cite{zuker_2}, a binomial form was proposed to describe the density of states for quantum many-body systems, such as those described by the nuclear shell model. For a system of dimension $N$, three parameters are required to define the binomial: $\cal N$ the effective dimension of the system, the asymmetry $p$, and an energy scale $\epsilon$. The span $S$ (the energy difference between the lowest and highest states), centroid $E_c$, variance $\sigma^2$, and dimensionless energy $x$ are given by
\begin{equation}
S = {\cal N}\epsilon,\hskip 0.15 cm E_c = {\cal N}p\epsilon,\hskip 0.15 cm \sigma^2 = {\cal N}pq\epsilon^2, \hskip 0.15 cm x =\frac{E}{S},
\end{equation}
where $p+q=1$ and obviously $E_c = H_1$ and $\sigma^2=M_2$. The binomial approximation to the level density is then given by
\begin{equation}
\label{eq:bin_ld}
\rho_b(x) = p^{x{\cal N}}q^{\bar x {\cal N}}
\frac{\Gamma({\cal N}+1)}{\Gamma(x{\cal N}+1)\Gamma(\bar x {\cal N}+1)}\frac{N{\cal N}}{S},
\end{equation}
with $\bar x = 1 - x$.
The binomial parameters $p$ and ${\cal N}$ can be determined by the 3$^{rd}$ and 4$^{th}$ moments of the Hamiltonian since for the binomial 
\begin{equation}
\label{eq:bin_m3}
R_3 = \frac{q-p}{\sqrt{{\cal N}pq}}
\end{equation}
and
\begin{equation}
\label{eq:bin_m4}
R_4 =  3 + \frac{1-6pq}{{\cal N}pq}.
\end{equation}
Defining $R = R_3^2/(R_4 - 3)$, the parameter $p$ becomes
\begin{equation}
\label{eq:bin_p}
p = \frac{1}{2}\left[ 1 - {\rm sgn}(M_3)\sqrt{1-2\left(\frac{1-R}{2-3R}\right)}~\right],
\end{equation}
from which, $\cal N$ follows directly from Eq.~(\ref{eq:bin_m4}). With $p$ and $\cal N$ known, the span is then given by
\begin{equation}
S = \sqrt{ \frac{{\cal N}\sigma^2}{pq}}.
\end{equation}
In addition, for the binomial, the ground-state energy is $E_{gs}^b = -Sp$, which may not correspond to the actual ground state energy $E_{gs}$. In this case, the level density in Eq.~(\ref{eq:bin_ld}) is shifted by $x - (E_c - Sp)/S$ so that the binomial centroid corresponds to the centroid of the Hamiltonian relative to the exact ground state. For the most part, the most significant hurdle in  implementing this approach has been the ability to compute $R_3$ and $R_4$, which can now be computed using the Lanczos method.

Note from Eq.~(\ref{eq:bin_p}), a real solution with $0 \le p \le 1$ requires $R \le 0$,  which implies $R_4 < 3$ and is representative of systems approaching an asymmetric Gaussian. Note that mathematically a solution for $p$ also exists when $R > 1$, which would imply $R_4> 3$ with a very large asymmetry. This solution, however, does not yield a physical solution where the $R_3$ and $R_4$ moments of the binomial correspond to the actual moments. Thus, the binomial is not applicable to the $^{76}$Ge results shown in Section~
\ref{sec:ELM}.

\begin{figure}
\includegraphics[scale=0.36]{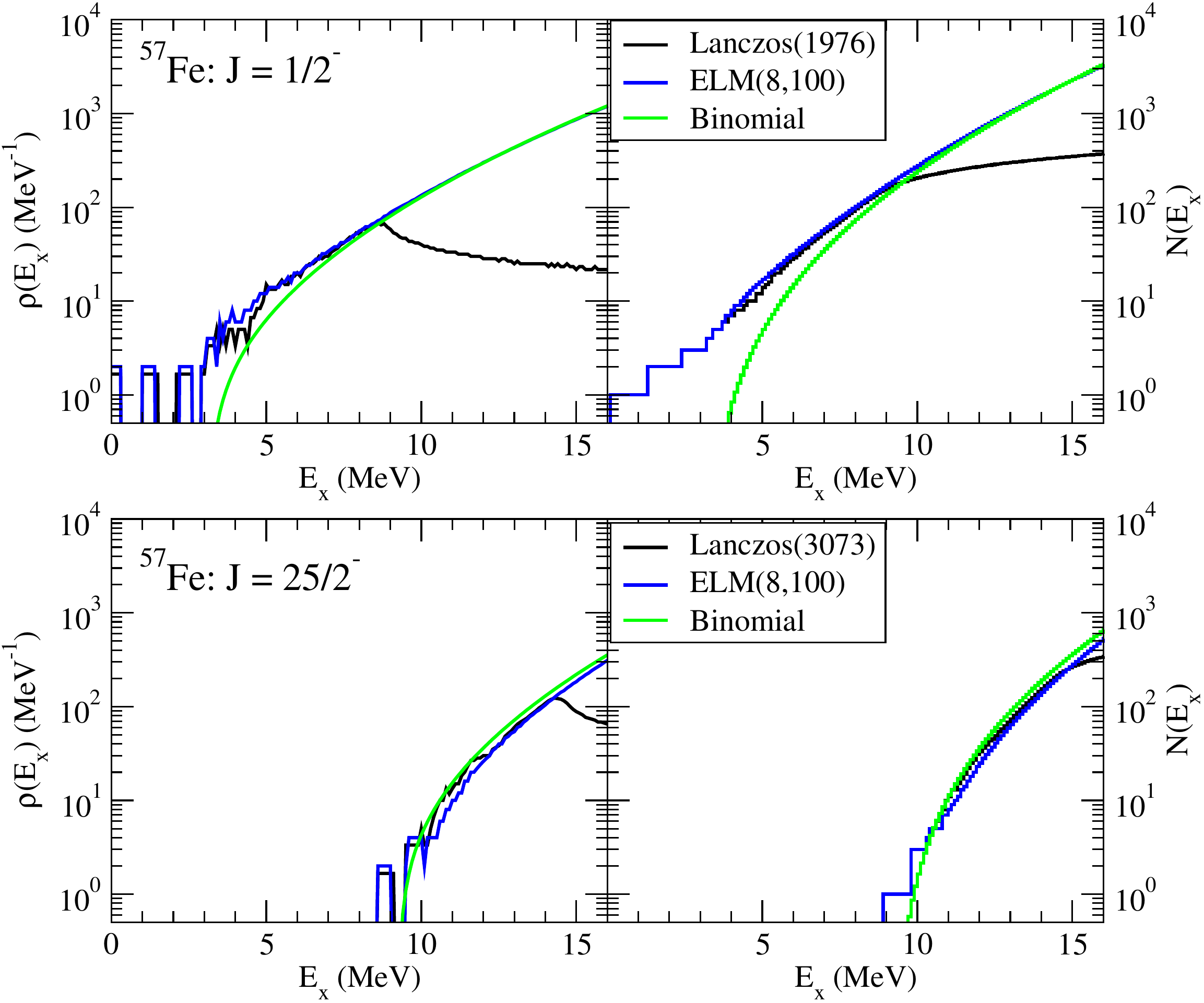}
\caption{
(color online) Results for the level density and cumulative density for $J^\pi=1/2^-$ and $25/2^-$states in $^{57}$Fe within the $1p0f$ shell-model space with the GXPF1A interaction. The black lines show the shell-model calculation, while the blue and green lines represent ELM(8,100) and the binomial approximation, respectively.}
\label{fig:ld_fe57_bin}
\end{figure}

In Figure~\ref{fig:ld_fe57_bin}, results for the level density and cumulative density for the $J^\pi = 1/2^-$ and $25/2^-$ states in $^{57}$Fe are shown for the the binomial approximation (green lines) and are compared to the ELM(8,100) (blue lines) and the shell model (black lines) obtained with a finite number of Lanczos iterations as specified in the figures. The figures show that both ELM and the binomial approximation are in agreement at higher excitation energies where the density of states is quite high. At lower excitation energies, the binomial approximation can be poor since it lacks information about the ground state and the low-lying spectrum, and in the case for the $J^\pi=1/2^-$ state in $^{57}$Fe, the ``effective'' lowest energy lies above the shell-model state.  This is not surprising since the binomial is limited to only four moments, and as was already pointed out, one would need of the order 40 moments (20 Lanczos iterations) for a reasonable calculation of the ground-state energy.  In addition, the low-energy behavior of the binomial is Gaussian-like, and thus, the level density tends to decrease dramatically at low energy, giving an effective lowest state so that $E_0^{\rm eff} > E_0$. For the most part, the ELM procedure can provide a better description of the low-lying spectrum if sufficient Lanczos iterations are performed in order to determine the energy of the lowest state in the specified model space. 

\subsection{Analytic Continuation of the Level Density}
\label{sec:AC}
As is shown in the previous sections, the level density modeled from the moments and shifted relative to the exact ground-state energy is a good representation of the exact shell-model level density at higher excitation energies. The principal question, however, is how to properly describe the level density in the situations illustrated in Figure~\ref{fig:ld_ib4k08} where the ELM has a discontinuity in the Lanczos matrix elements and Figure~\ref{fig:ld_fe57_bin} where the binomial approximation substantially undershoots the shell-model result. In both cases, the moments by themselves dramatically miss the lowest energy, $E_0$ in the configuration space, leading to an ``effective'' $E_0^{\rm eff}$ that is too high in energy. In principle, the ELM(8,$N_{\rm Lanc}$) procedure will work by ensuring that the modeled and exact Lanczos matrix elements are reasonably matched so that there isn't a discontinuity larger than natural noise in the calculated matrix elements. In some cases, however, the number of Lanczos iterations, $N_{\rm Lanc}$ required would be prohibitively large, which in effect negates any advantages in the approach. 

A strategy for the case when $E_0^{\rm eff} > E_0$ is similar to that outlined by Gilbert and Cameron~\cite{Gilbert-Cameron} where the goal was to describe the level density via two components: an exponentially increasing function at low energy that is then matched to the back-shifted Fermi gas at higher energies. Here, we take a similar approach by matching an exponentially increasing level density to the ELM level density at a matching energy $E_m$. Thus, at low energy, the density of states is taken to be
\begin{equation}
\label{eq:exp_rho}
\rho(E_x) = \exp \left[ (E_x - E_{\rm shift})/T\right].
\end{equation}
Note that $E_{\rm shift}$ specifies that the for cumulative density we have $N(E_{\rm shift})=1$. The exponential level density of Eq.~(\ref{eq:exp_rho}) can then be matched at energy $E_m$ to the ELM or binomial approximation by requiring continuity in the level density and by defining the temperature $T$ as the inverse of the logarithmic derivative of $\rho$, i.e.,
\begin{equation}
\label{eq:temp}
T(E_x) = \frac{\rho(E_x)}{\rho'(E_x)}.
\end{equation}
At a given $E_m$, the continuity requirement for the level density specifies $E_{\rm shift}$ as
\begin{equation}
\label{eq:Ematch}
E_{\rm shift}(E_m) = E_m - T(E_m)\ln\left[ \rho(E_m)\right].
\end{equation}
Thus, the matching energy can be chosen so that $E_{\rm shift}(E_m) = E_0$. Practical considerations for finding the matching energy for the ELM procedure are given in Appendix~\ref{app:B}

\begin{figure}
\includegraphics[scale=0.36]{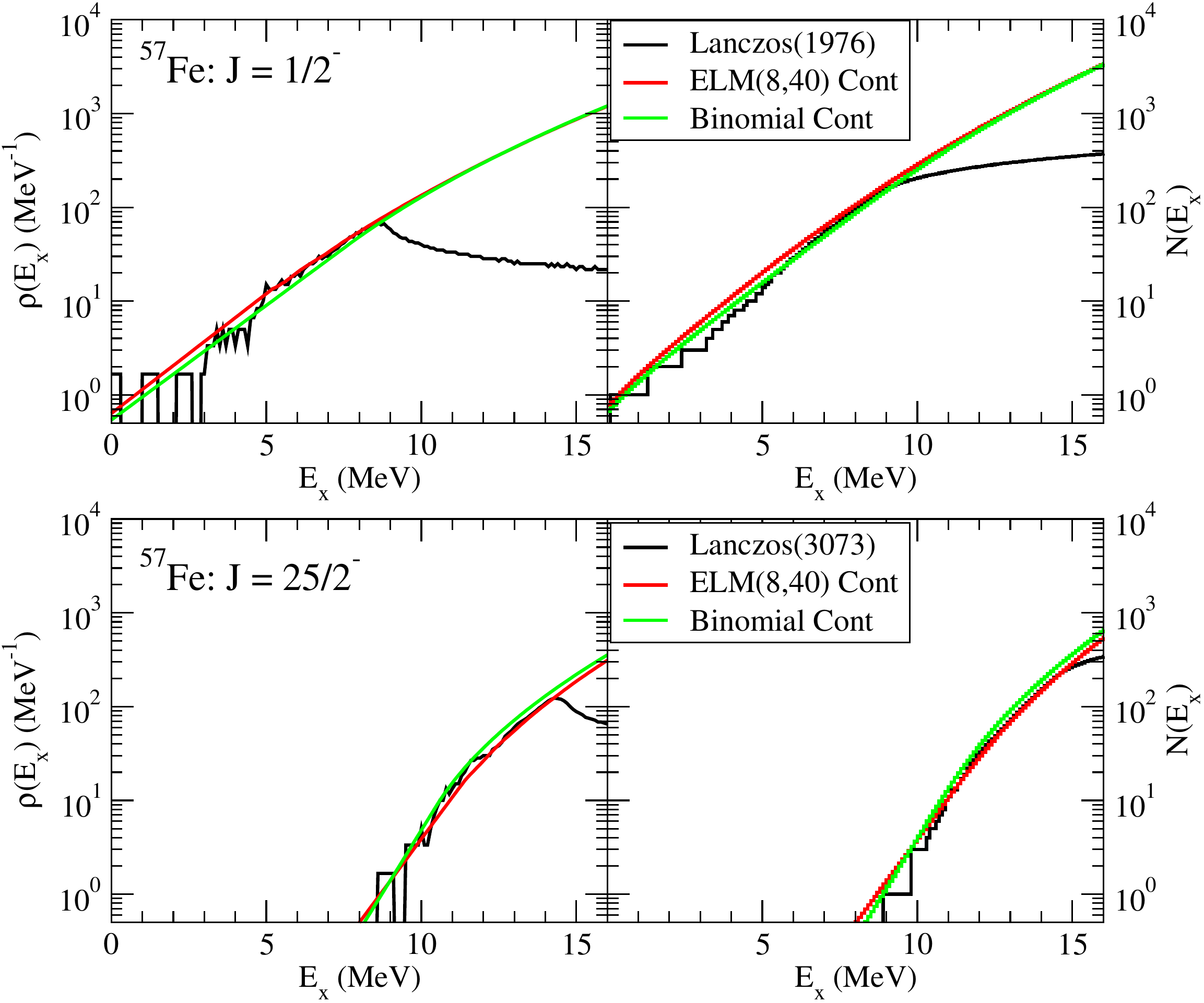}
\caption{
(color online) Results for the level density and cumulative density for $J^\pi=1/2^-$ and $25/2^-$ states in $^{57}$Fe within the $1p0f$ shell-model space with the GXPF1A interaction. The black lines show the shell-model calculation, while the red and green lines represent the ELM$_{\rm AC}$(8,40) and binomial calculations, as described in the text, respectively.}
\label{fig:ld_fe57_cont}
\end{figure}

\begin{figure}
\includegraphics[scale=0.36]{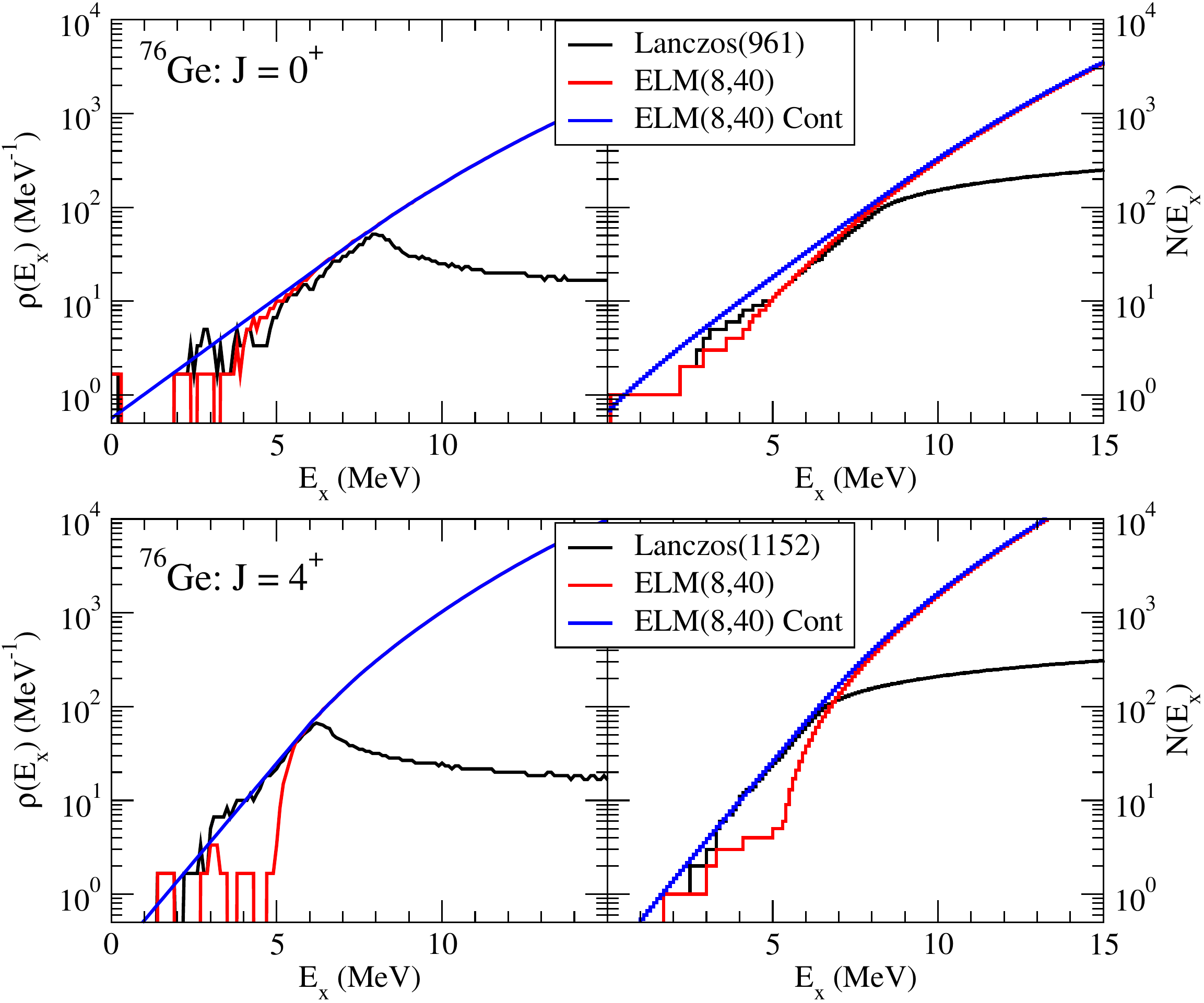}
\caption{
(color online) Results for the level density and cumulative density for $J^\pi=0^+$ and $4^+$ states in $^{76}$Ge within the $jj44$ shell-model space with the $jj44b$ interaction. The black lines show the shell-model calculation, while the red and blue lines represent ELM(8,40) and the ELM$_{\rm AC}$(8,40) as described in the text, respectively.}
\label{fig:ld_ge76_cont}
\end{figure}

In Figure~\ref{fig:ld_fe57_cont}, results for the ELM analytic continuation, ELM$_{\rm AC}$(8,40), and the binomial level densities are shown for the $J^\pi = 1/2^-$ and $25/2^-$ states in $^{57}$Fe, while the ELM$_{\rm AC}$(8,40) level density for the $J^\pi=0^+$ and $4^+$ states in $^{76}$Ge are shown in Figure~\ref{fig:ld_ge76_cont} (note that the binomial approach is not applicable due to $R_4 > 3$). Overall, the extrapolation works well; especially when the effective lowest state for the modeled level density is higher than the actual, i.e., $E_0^{\rm eff} > E_0$. Under this condition, it is possible to smoothly match the modeled level density down to the lowest state. As can be seen in Figures~\ref{fig:ld_fe57_cont} and  \ref{fig:ld_ge76_cont}, however, in some cases, such as the lower spin, the extrapolated level density tends to miss a ``gap'' in the excitation spectrum at low excitation energies. This most likely reflects the effect of pairing. 

\begin{figure}
\includegraphics[scale=0.37]{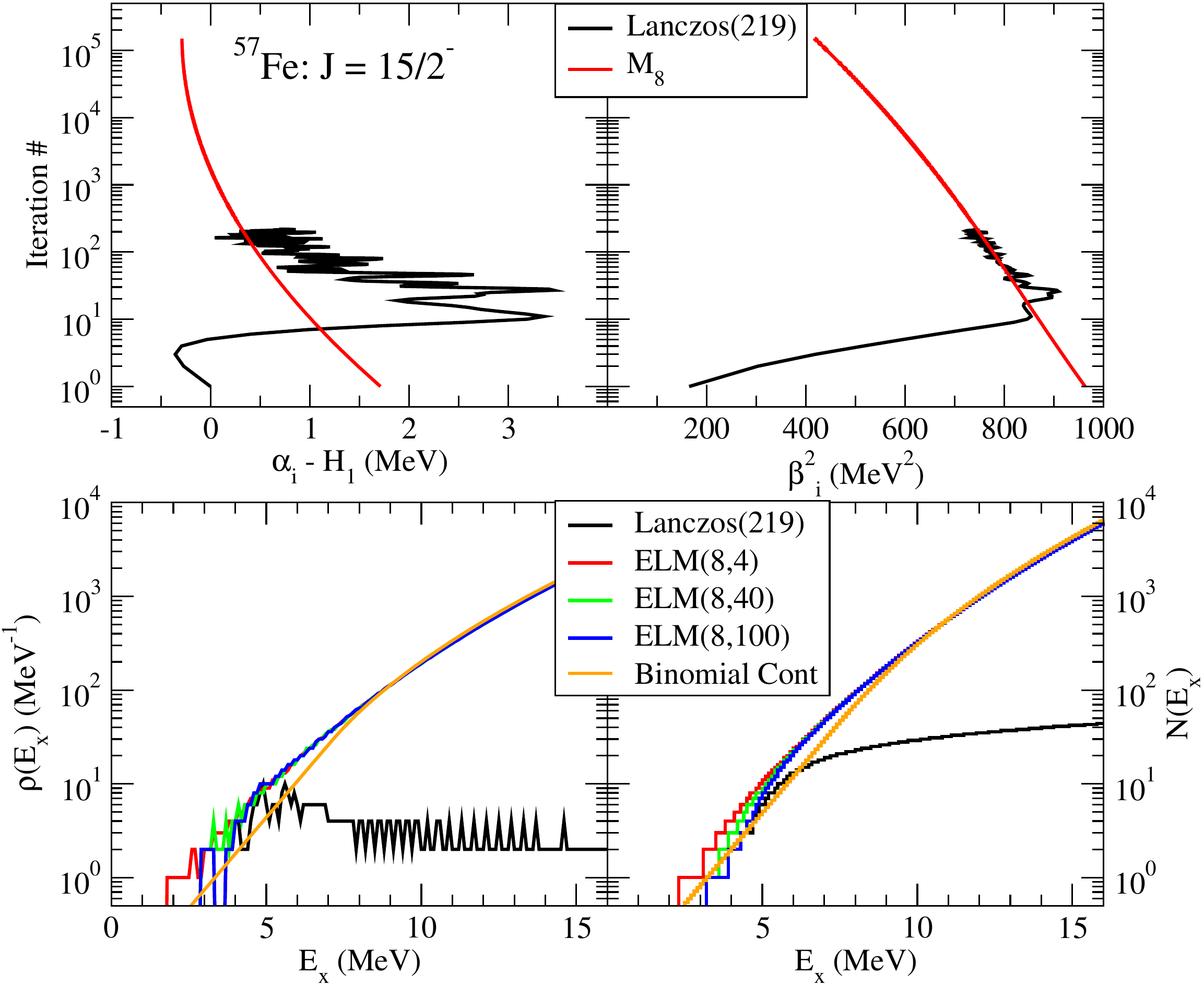}
\caption{
(color online) Comparison of results with ELM(8,$N_{\rm Lanc}$ for $N_{\rm Lanc} = 4$ (red), $40$ (green) , and $100$ (blue) for the $J^\pi = 15/2^-$ space in $^{57}$Fe within the $1p0f$-shell model space with the GXPF1A interaction. In addition the shell model with 219 Lanczos iterations (black) and the analytically continued binomial (orange) are also shown.}
\label{fig:ld_fe57_elm}
\end{figure}

The case where $E_0^{\rm eff} < E_0$ is less common and is generally not possible with the binomial level density due to the high curvature of the Gaussian, which tends to decrease the level density dramatically at low excitation energy. However, this can occur for the ELM when a small number of actual Lanczos iterations, $N_{\rm Lanc}$, is used. On the other hand, for ELM, better agreement with the low-lying spectrum is achieved with increasing $N_{\rm Lanc}$, which is also needed in order to obtain a reasonable estimate of $E_0$. Shown in Figure~\ref{fig:ld_fe57_elm} is the case of the $J^\pi = 15/2^-$ space in $^{57}$Fe (within the $1p0f$-shell model space with the GXPF1A interaction), where results for ELM(8,4) (red), ELM(8,40) (green), and ELM(8,100) (blue) are shown in comparison to the shell model with 219 Lanczos iterations (black) and the analytically continued binomial approximation (orange). All the modeled level densities are in agreement at high energy, where the spectrum is dominated by the statistical properties of the Hamiltonian. The agreement between ELM(8,$N_{\rm Lanc}$) and the shell model at low excitation energy improves with increasing $N_{\rm Lanc}$ as is to be expected. Indeed, reasonable agreement is achieved with ELM(8,40) which is close to the minimum number of iterations needed to give an accurate energy for $E_0$ and the next level. 

\begin{figure}
\includegraphics[scale=0.47]{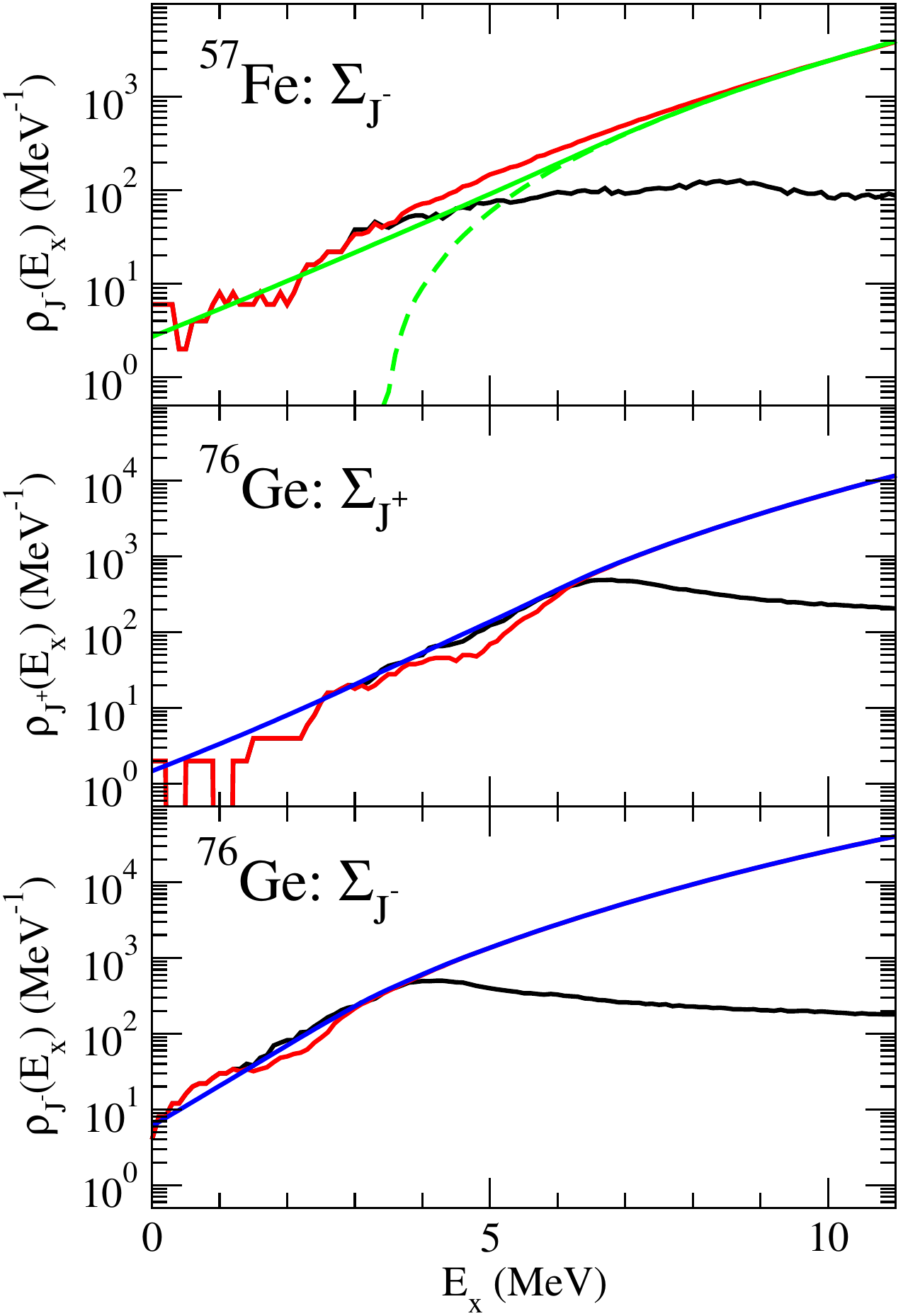}
\caption{
(color online) Comparison of results of the level density summing all angular momenta of a given parity for $^{57}$Fe ($\sum_{J^-}$), $^{76}$Ge ($\sum_{J^+}$), and $^{76}$Ge ($\sum_{J^-}$). The black lines are from the Lanczos iterations, the red line is the ELM(8,100) reconstruction, the blue and green lines are ELM$_{\rm AC}$(8,40) results. The dashed green line is from the binomial, while the green line is the analytic continuation of the binomial. ($^{57}$Fe only). }
\label{fig:ld_rho_sum}
\end{figure}

Shown in Figure~\ref{fig:ld_rho_sum} are results for the various approaches for the summed over angular momenta with fixed parity. The black lines show the results from the Lanczos iterations while the red lines are the ELM(8,100) results. The blue lines show the ELM$_{\rm AC}$(8,40) results. The green line is the result for the analytically continued binomial, while the dashed green line is the binomial ($^{57}$Fe only).  For the most part, both the ELM and binomial agree at high excitation energy. In general, the most successful approach is ELM(8,$N_{\rm Lanc}$) where $N_{\rm Lanc}$ is large enough to capture key features of the Lanczos matrix elements. As discussed previously, this is when the difference between the modeled and actual Lanczos matrix elements is less than the natural ``noise'' in the matrix elements; that is no strong discontinuities. For $^{57}$Fe, this is generally achieved with $N_{\rm Lanc} \approx 50 - 100$. In this sense, the ELM(8,100) results are likely representative of the full shell model with 200,000 iterations. For $^{57}$Fe, analytically continuing the binomial to $E_0$ is a significant improvement over the binomial itself. It does, however, tend to underestimate the actual level density in the region $E_x \approx 3 - 8$ MeV. For $^{76}$Ge, one would need $N_{\rm Lanc} \ge 1000$ in order to avoid the most significant discontinuities. A situation that is less than ideal. On the other hand, analytically continuing the ELM(8,40) gives a good overall description of the level density.

\section{Applications of the ELM: $^{57}$Fe, $^{74}$Ge, and $^{76}$Ge}

\begin{figure}
\centering
\scalebox{0.5}{\includegraphics{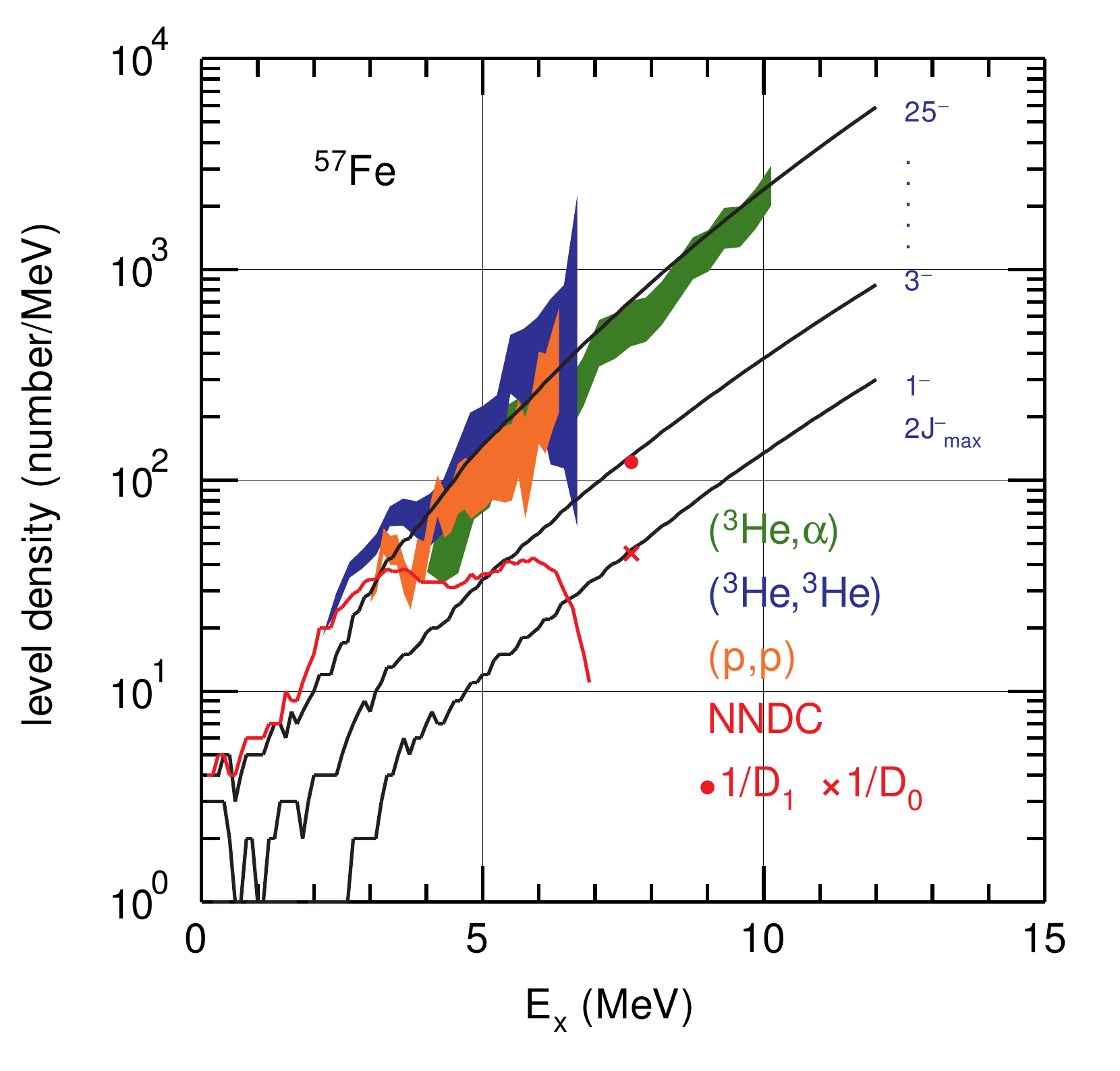}}
\caption{
(color online) Level densities for $^{57}$Fe. The black lines show the calculated angular-momentum summed level density for negative-parity states up to the value of $2J^\pi_{\rm max}$ as indicated to the right of each line. The experimental $\ell=0$ and $1$ level densities~\cite{mug} at $E_x = S_n$ are shown by the red cross and circle, respectively, with error bars about the size of the symbols. Note that the calculated level densities are only for the negative parity states contained within the  $1p0f$ shell model space. Other data is for the sum of both negative and positive parity states. The red line shows the experimental level density obtained from the states listed in NNDC~\cite{nndc}. Level densities inferred from reaction data are shown by the shaded areas: (green) $^{55}$Mn($^{3}$He,$\alpha$) reaction \cite{Voinov}, (blue) $^{57}$Fe($^{3}$He,$^{3}$He$^\prime$) reaction \cite{Algin}, and (orange) $^{57}$Ni(p,p$^\prime$) reaction \cite{Lar17}.}
\label{fig:ld57}
\end{figure}

We now apply the extrapolated Lanczos method to compute the level density for $^{57}$Fe within the $0p1f$ model space using the GXPF1A interaction. The level densities for each negative parity, angular momentum configuration space were computed with ELM(8,100) and are shown in Figure~\ref{fig:ld57}. The black lines show the angular-momentum summed level density for negative parity states up to the $2J^\pi_{\rm max}$ value indicated to the right of each line. The experimental $\ell$=1 level density~\cite{mug} ($\rho_{1/2^-}$ + $\rho_{3/2^-}$) at $  E_{x}=S_{n}  $ (the neutron decay threshold) is shown with the red circle (the error bar is approximately equal to the size of the circle). The experimental value for $\ell$=0 level density~\cite{mug} ($\rho_{1/2^-}$) at $E_x=S_n$ is shown with the red cross (the error bar is approximately equal to the size of the cross). Other data shown in the figure is for the sum of both positive and negative parity states. The red line shows the level density obtained from the experimentally observed states listed in the NNDC~\cite{nndc}. The shaded areas are the bounds inferred from the various reaction data (see Figure caption)~\cite{Lar17,Algin,Voinov}. We note that the $1p0f$ shell model space does not contain any positive parity states for $^{57}$Fe.

The agreement between our calculation (the sum of densities for states with 1/2$^{-}$ and 3/2$^{-}$) and the $\ell$=1 level density is excellent. In addition, the level density for $1/2^+$ states is nearly the same as that computed for $1/2^-$ states, which indicates that the parity ratio is close to unity at $E_x = S_n$. Thus, our estimate of the total level density would be a factor of two larger than shown in Fig.~\ref{fig:ld57}. The level density obtained from NNDC~\cite{nndc} levels (see Figure~\ref{fig:ld57}) becomes about a factor of two larger than that  calculated for negative parity states starting  around $E_{x}$ = 3 MeV, indicating a parity ratio close to unity around 3 MeV. Taking this into account, the total level density above 3 MeV should be a factor of two larger than that for negative parity states alone.  The overall agreement between the calculated level density and that inferred from reaction data is reasonable. However, the differences exhibited between the different reactions and the fact that the inferred level densities are of the order as those computed here suggest that each reaction might be more selective than expected and the analysis is potentially missing states.

A proper treatment of the 1/2$^{ + }$ level density for Fe nuclei must take into account particle-hole excitations beyond the $1p0f$ model space. For example, for $^{57}$Fe we should consider the coupling of the $  \nu  (0g_{9/2})  $ particle orbital to the calculated level density of (4,5)$^{ + }$ states of $^{56}$Fe, and the coupling of $  \pi  (0d_{3/2},1s_{1/2})  $ hole orbitals to the calculated level density of (0,1,2,3)$^{ + }$ states of $^{58}$Co. This extension will be explored in the future.

\begin{figure}
\centering
\scalebox{0.6}{\includegraphics{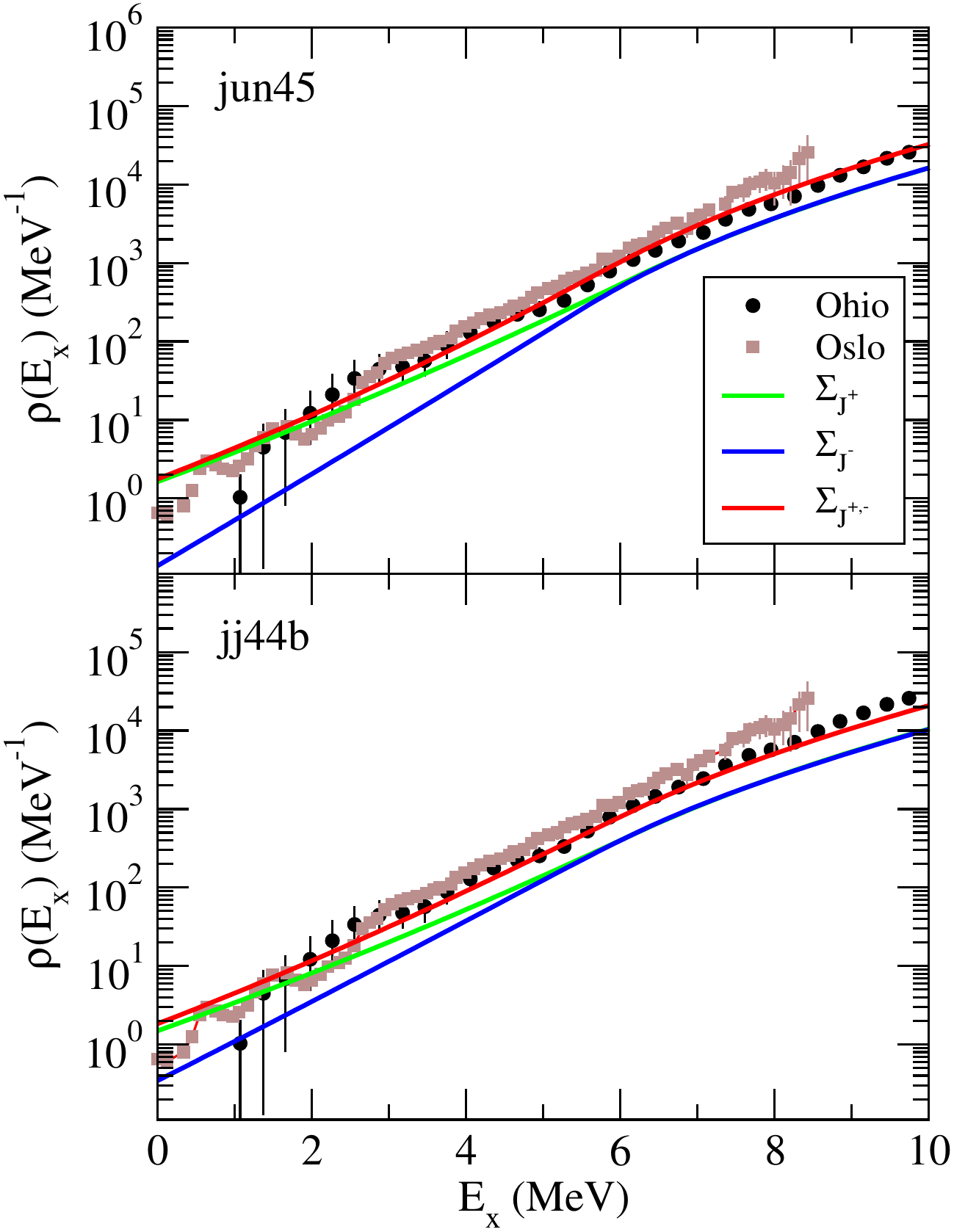}}
\caption{
(color online) Level densities for $^{74}$Ge compared with experimental values. The black points, labeled Ohio, are inferred from proton evaporation spectra~\cite{Voinov-2}, while the brown squares, labeled Oslo, are from the Oslo method~\cite{Ge74-Oslo}. Level densities are shown for two shell model interactions, jun45 (upper) and $jj44b$ (lower). The green and blue lines represent the total level density for positive- and negative-parity states, respectively, while the red line is the total level density. }
\label{fig:ld74ge}
\end{figure}

\begin{figure}
\centering
\scalebox{0.6}{\includegraphics{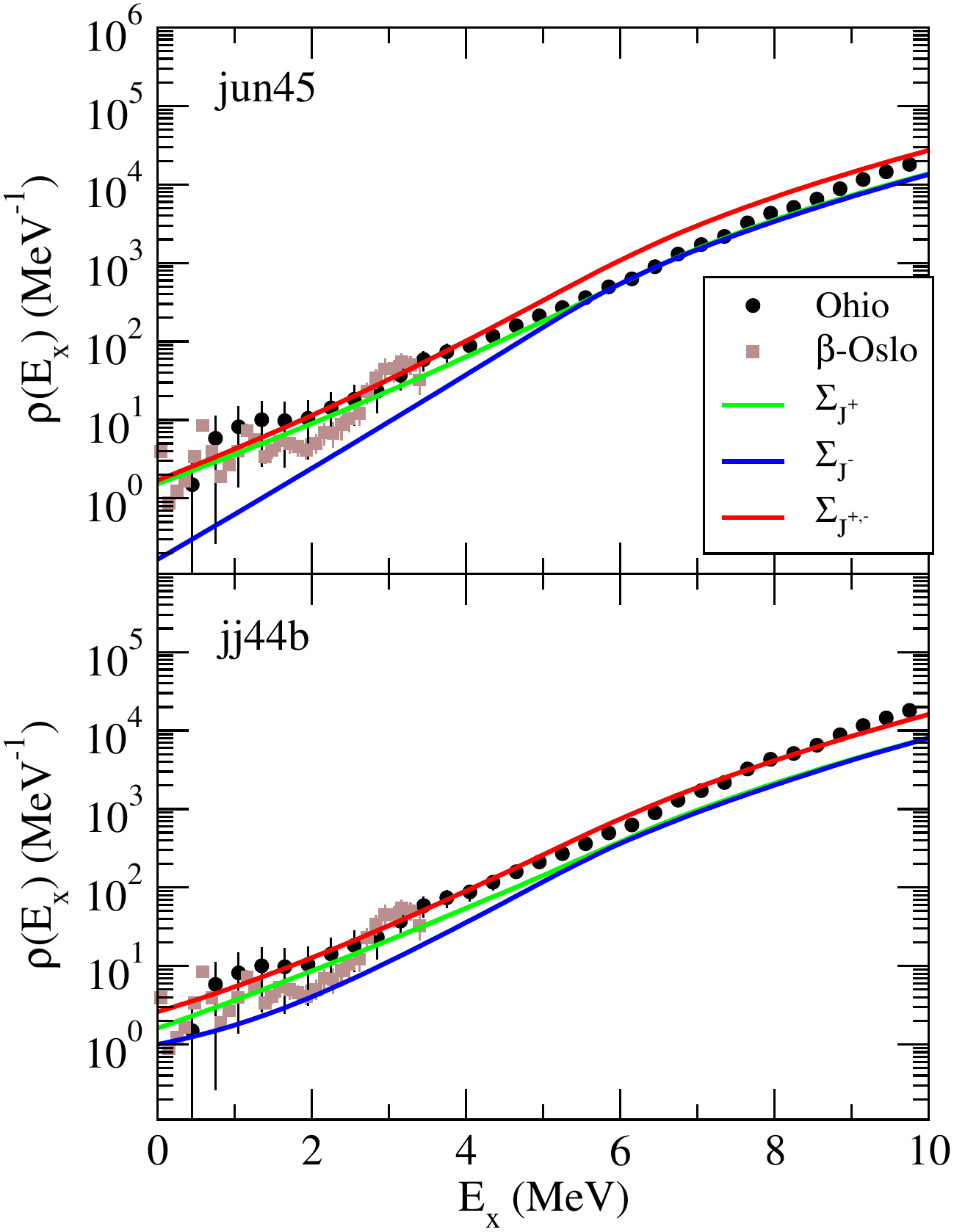}}
\caption{
(color online) Level densities for $^{76}$Ge compared with experimental values. The black points, labeled Ohio, are inferred from proton evaporation spectra~\cite{Voinov-2}, while the brown squares, labeled Oslo, are from the $\beta$-Oslo method~\cite{Ge76-Oslo}. Level densities are shown for two shell model interactions, jun45 (upper) and $jj44b$ (lower). The green and blue lines represent the total level density for positive- and negative-parity states, respectively, while the red line is the total level density.}
\label{fig:ld76ge}
\end{figure}

In Figures~\ref{fig:ld74ge} and \ref{fig:ld76ge}, the ELM$_{\rm AC}$(8,100) results are shown for the nuclei $^{74}$Ge and $^{76}$Ge within the $jj44$ shell-model space and the $jj44b$ and jun45 interactions in comparison with experimental values inferred from proton evaporation spectra resulting from the compound nuclear reactions $^{68,70}$Zn($^7$Li,Xp) (black circles)~\cite{Voinov-2}. In addition, for $^{74}$Ge, results~\cite{Ge74-Oslo} from the Oslo method are shown (brown squares), while for $^{76}$Ge, results~\cite{Ge76-Oslo} from the $\beta$-Oslo method are shown. Note that the Oslo method requires a normalization, which was extracted from the experimental $D_0$ value. Overall, the agreement between the ELM$_{\rm AC}$(8,100) results and those inferred from proton-evaporation spectra are excellent up to $E_x\approx 8-9$~MeV. This is well within the expectation that the shell model provides an accurate representation of the excitation spectrum up to the point where intruder states appear.

\begin{table}
\caption{Comparison between calculated and experimental~\cite{mug} level spacings for $\l=1$ neutron resonances ($D_1$) for various Fe isotopes. The neutron separation energy, $S_n$, for the isotope of listed and the angular momentum $J_t^\pi$ for the target $^{A-1}$Fe nucleus are shown.}
\begin{ruledtabular}
\begin{tabular}{rrrrr}
 & $J_t^\pi$ & $S_n$ (MeV) & $D_1^{\rm calc}$ (keV)  & $D_1^{\rm exp}$ (keV)\\
 \hline
$^{55}$Fe & $0^+$              &  9.298  & 5.6 & 4.75$\pm$0.15 \\
$^{57}$Fe & $0^+$              &  7.646  & 7.6 & 8.21$\pm$0.48 \\
$^{58}$Fe & $\frac{1}{2}^-$ & 10.044 & 3.3 & 2.58$\pm$0.26 \\
$^{59}$Fe & $0^+$              &  9.298  & 11.6 & 5.03$\pm$0.30 \\
\end{tabular}
\end{ruledtabular}
\label{tab:Fe_D1}
\end{table}

\begin{table}
\caption{Comparison between experimental~\cite{mug} and calculated (with the $jj44b$ and jun45 interactions) level spacings for $\l=0$ neutron resonances ($D_0$) for various Ge isotopes. The neutron separation energy, $S_n$, for the isotope of listed and the angular momentum $J_t^\pi$ for the target $^{A-1}$Ge nucleus are shown.}
\begin{ruledtabular}
\begin{tabular}{rrrrrr}
 & $J_t^\pi$ & $S_n$ (MeV) & $D_0^{\rm calc}$ (keV)  & $D_0^{\rm calc}$ (keV)  & $D_0^{\rm exp}$ (keV)\\
 &                &                       &     $jj44b$                          &   jun45                            &      \\
 \hline
$^{73}$Ge & $0^+$                 &  6.782    & 6.6    & 4.3   & 2.07$\pm$0.29 \\
$^{74}$Ge & $\frac{9}{2}^+$   &  10.196  & 0.33  & 0.23  & 0.099$\pm$0.001 \\
$^{75}$Ge & $0^+$                 &   6.505   & 8.9    &  5.5   & 3.0$\pm$1.5 \\
$^{77}$Ge & $0^+$                 &    6.076  & 18.14 & 10.6  & 4.82$\pm$0.76 \\
\end{tabular}
\end{ruledtabular}
\label{tab:Ge_D0}
\end{table}

To conclude this section, calculated values for the level spacings for Fe and Ge isotopes are shown in Tables~\ref{tab:Fe_D1} and \ref{tab:Ge_D0}, respectively. For Fe isotopes, level spacing for $l=1$ neutron resonances, $D_1$, are shown, while for Ge isotopes, the level spacings for $l=0$ neutron resonances are displayed. The experimental neutron separation energy, $S_n$, which is equivalent to the excitation energy of the system of interest, is tabulated as well as the angular momentum and parity, $J_t^\pi$, of the target ${A-1}$ nucleus. The experimental data are from Ref.~\cite{mug}. For the Ge isotopes, results are shown for the two shell-model Hamiltonians $jj44b$ and jun45. Overall, good agreement is achieved for Fe isotopes except for $^{59}$Fe, which is likely signaling an increasing importance of the $0g_{9/2}$ orbit as more neutrons are added. For the Ge isotopes, the calculated $D_0$ values are larger than experiment. This implies that the computed level densities are too small, which is in contradiction with the agreement with the level densities inferred from proton-evaporation spectra as shown in Figs.~\ref{fig:ld74ge} and \ref{fig:ld76ge}. The jun45 interaction has a larger level density and generally yields a $D_0$ value within a factor of two from experiment. The exception is $^{74}$Ge, but here $S_n = 10.196$~MeV, which from Fig.~\ref{fig:ld74ge}, is an excitation energy about 1-2 MeV above where the model space is valid. On the other hand, we note the overall good agreement between our Ge calculations and the data from Ref.~\cite{Voinov-2} shown in Figs.~\ref{fig:ld74ge} and \ref{fig:ld76ge}.

\section{Angular Momentum Dependence of the Level Density}

\begin{figure}
\centering
\scalebox{0.3}{\includegraphics{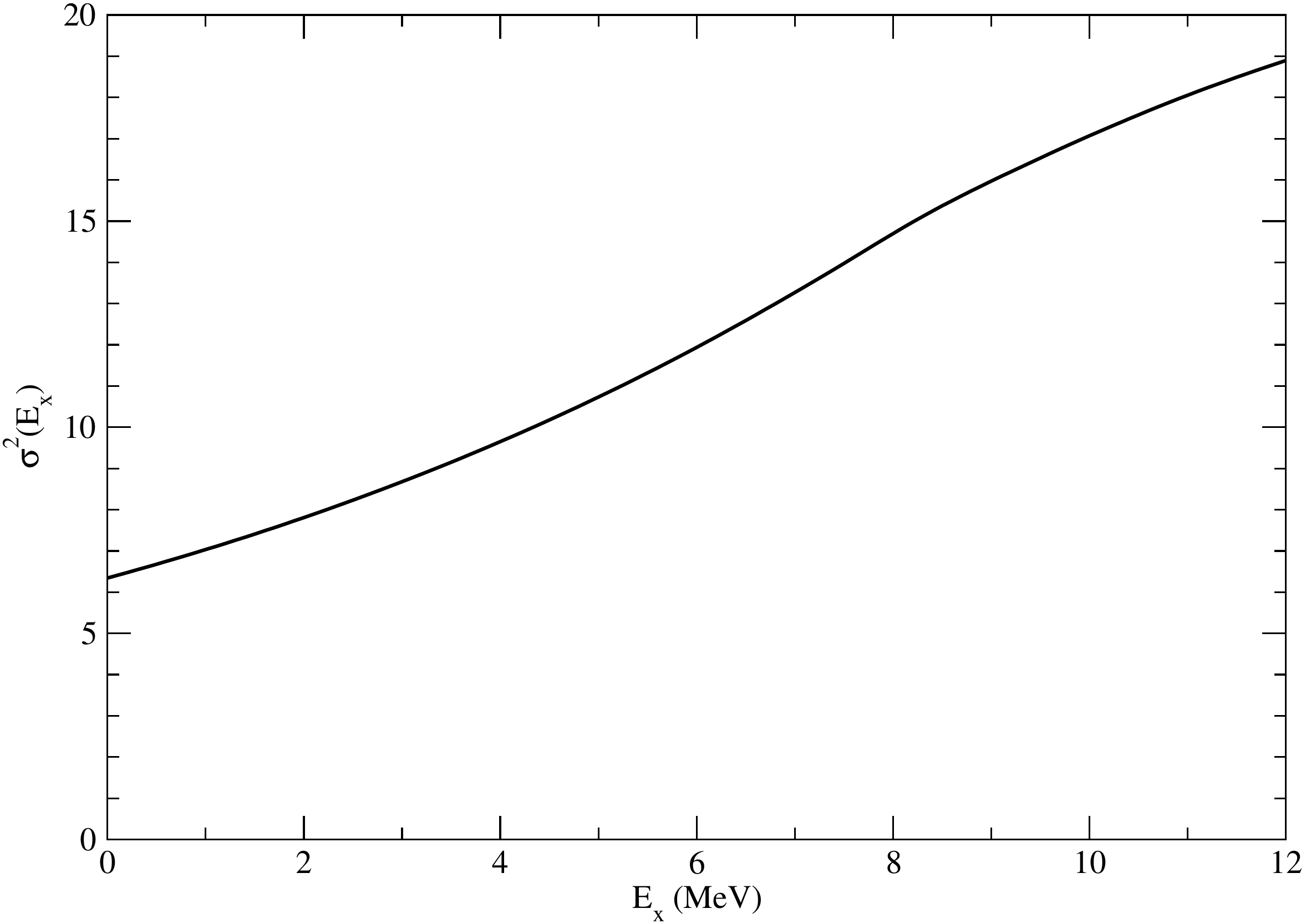}}
\caption{Calculated spin cutoff parameter for $^{57}$Fe as a function of excitation energy.}
\label{fig:Fe57-spincut}
\end{figure}

\begin{figure}
\centering
\scalebox{0.4}{\includegraphics{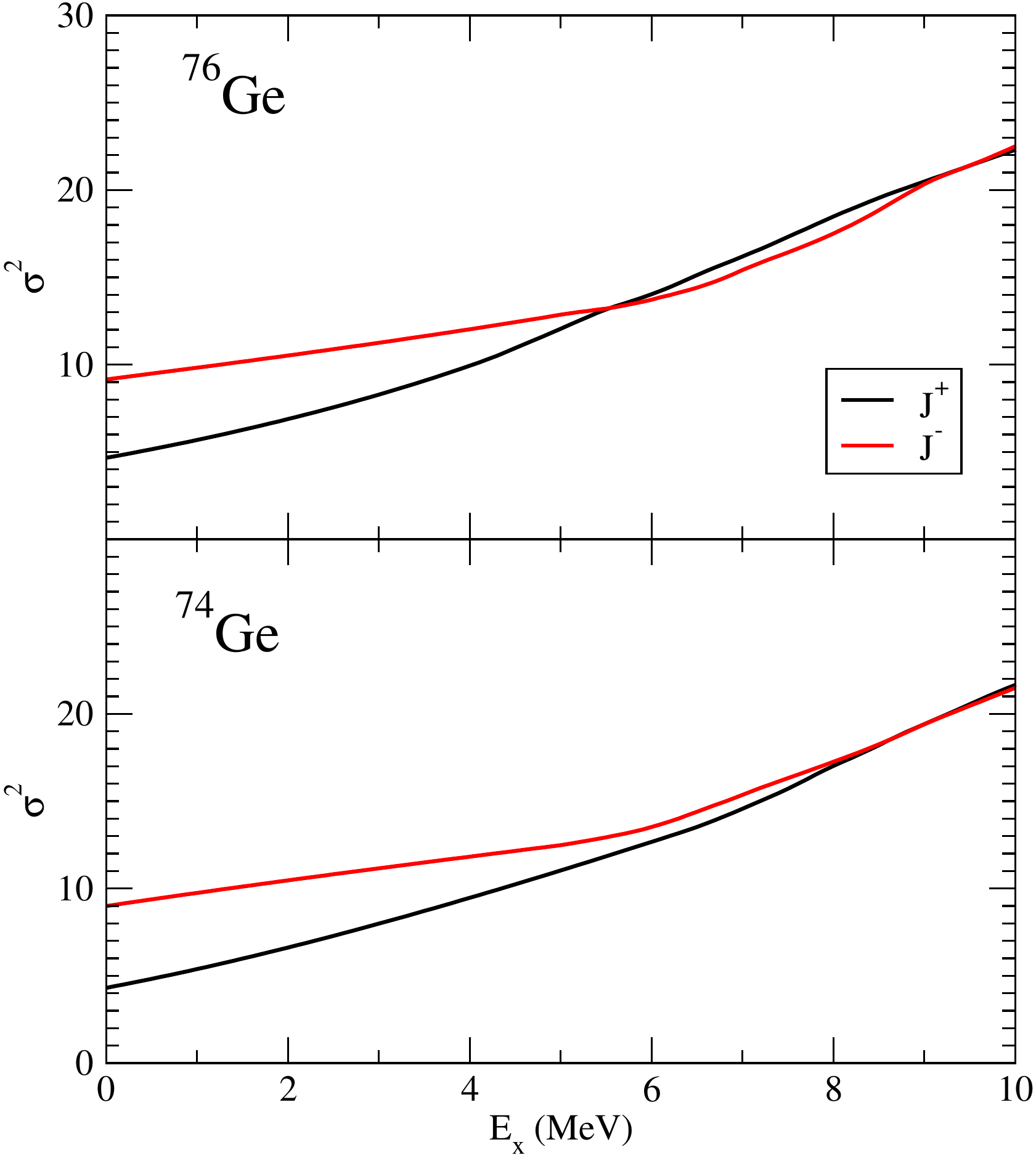}}
\caption{Calculated spin cutoff parameter for $^{74}$Ge  and $^{76}$Ge as a function of excitation energy. The black and red lines represent the positive- and negative-parity spaces, respectively. }
\label{fig:Ge-spincut}
\end{figure}

The angular momentum dependence of the level density is key to understanding many reactions.  A commonly used form comes from the original work of Bethe \cite{bethe}, where the level density for a given $J$ is 
\begin{equation}
\rho(E_x,J) = P(J) \rho(E_x)
\end{equation}
with
\begin{equation}
P(J) = \frac{(2J+1)}{2\sigma ^{2}} \,\, {\rm exp}\left [-(J+1/2)^{2}/2\sigma^{2} \right ],
\label{eq:PJ}
\end{equation}
and $\sigma^{2}$ being the so-called spin cutoff parameter, which is energy dependent. The spin cutoff parameter can be determined at a fixed excitation energy via
\begin{equation}
\sigma ^{2} = \langle (J+1/2)^{2}\rangle /2.
\label{eq:spincut}
\end{equation}
The calculated spin cutoff parameters for $^{57}$Fe, $^{74}$Ge, and $^{76}$Ge as a function of excitation energy are shown in Figures~\ref{fig:Fe57-spincut} and \ref{fig:Ge-spincut}. In Figure~\ref{fig:Ge-spincut}, both the positive- and negative-parity spin cutoff parameters are shown. 

\begin{figure}
\centering
\scalebox{0.45}{\includegraphics{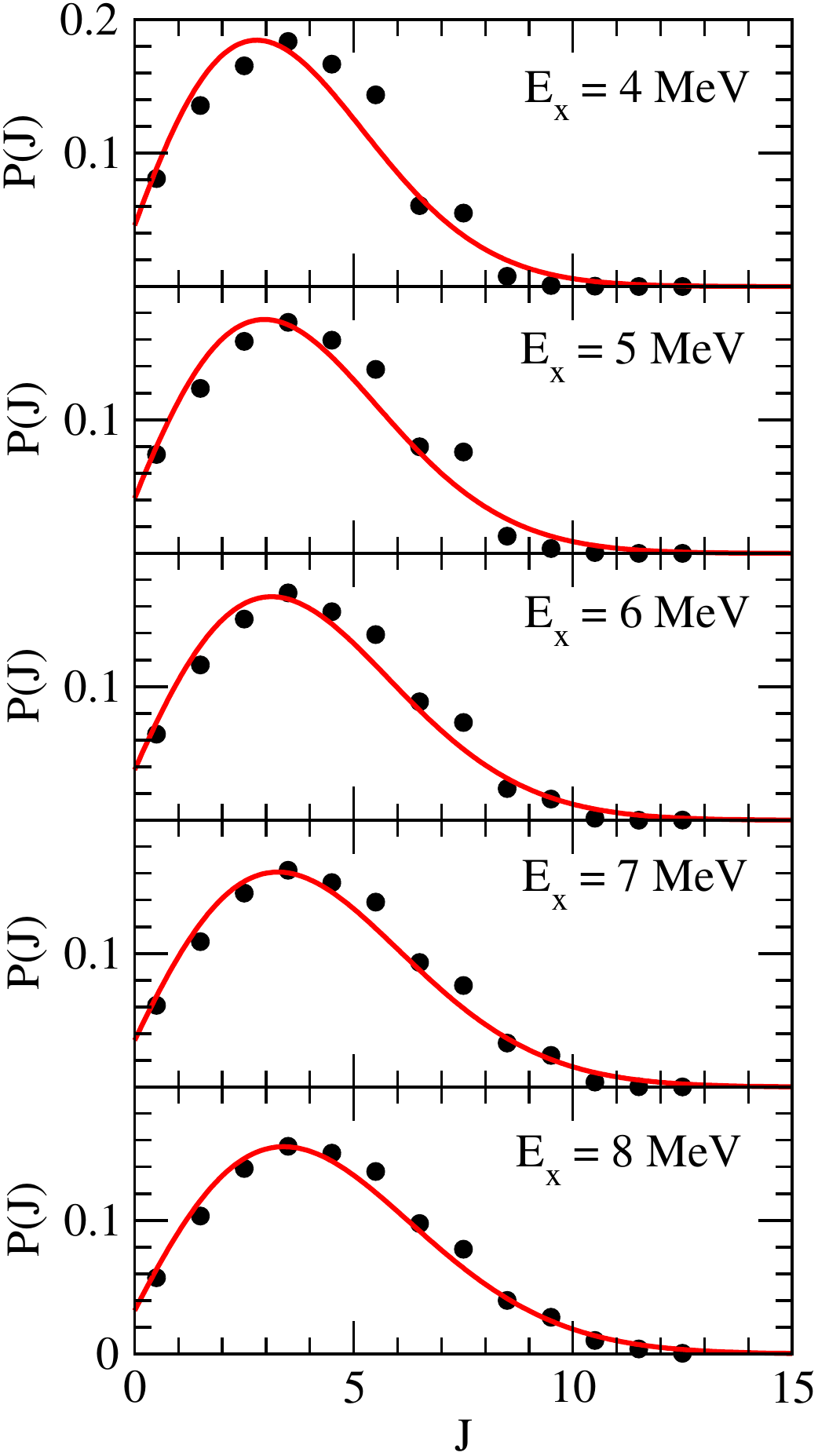}}
\caption{Angular momenta probabilities for $^{57}$Fe are shown for five excitation-energy slices across the density of states. The red lines show the results obtained from Eq.~(\ref{eq:PJ}) with the spin cutoff parameter shown in Figure~\ref{fig:Fe57-spincut}.}
\label{fig:Fe57-Jsq}
\end{figure}

\begin{figure}
\centering
\scalebox{0.5}{\includegraphics{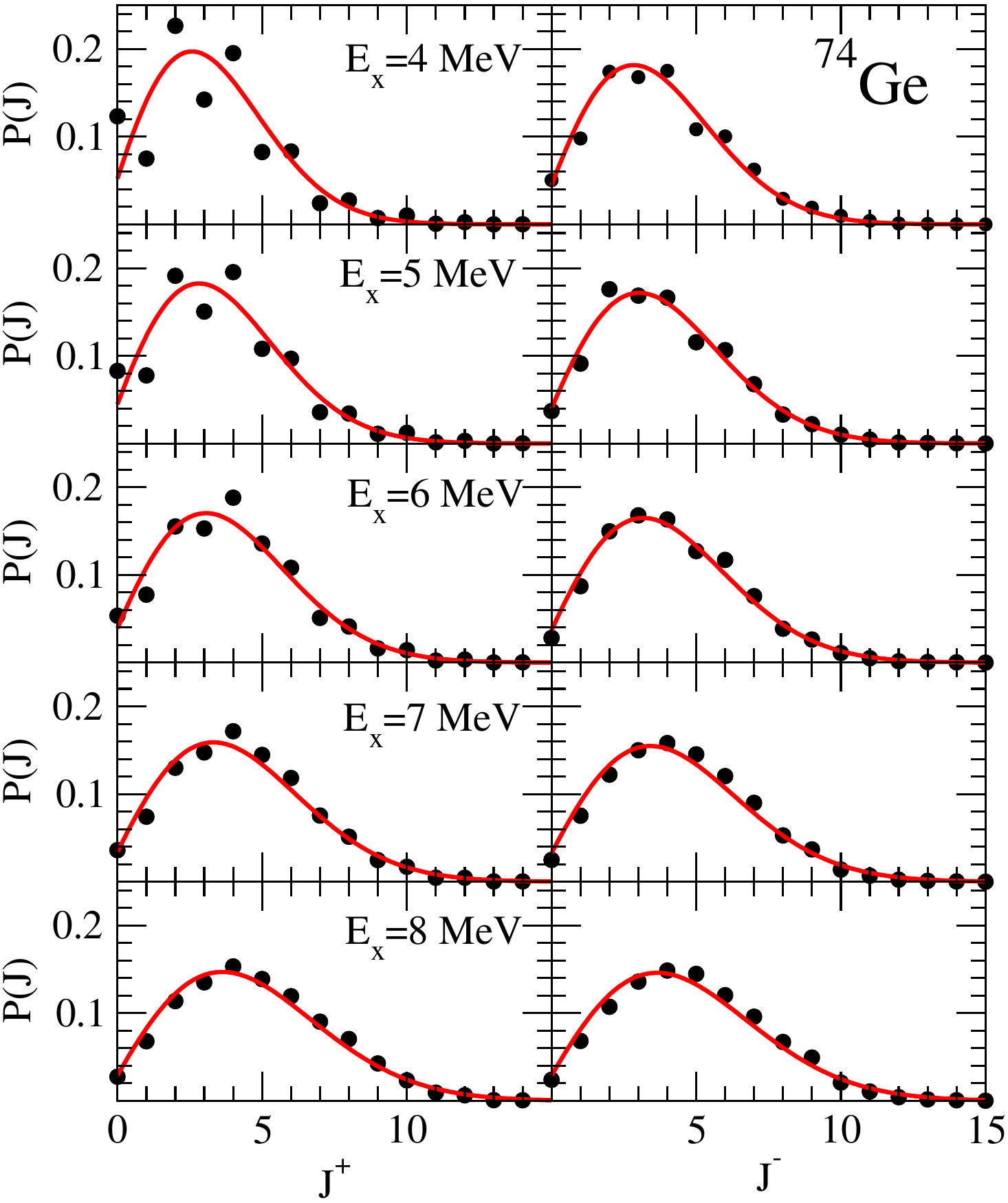}}
\caption{Angular momenta probabilities for $^{74}$Ge are shown for five excitation energies across the positive- and negative-parity states. The red lines show the results obtained from Eq.~(\ref{eq:PJ})  with the spin cutoff parameter shown in Figure~\ref{fig:Ge-spincut}.}
\label{fig:Ge74-Jsq}
\end{figure}

\begin{figure}
\centering
\scalebox{0.5}{\includegraphics{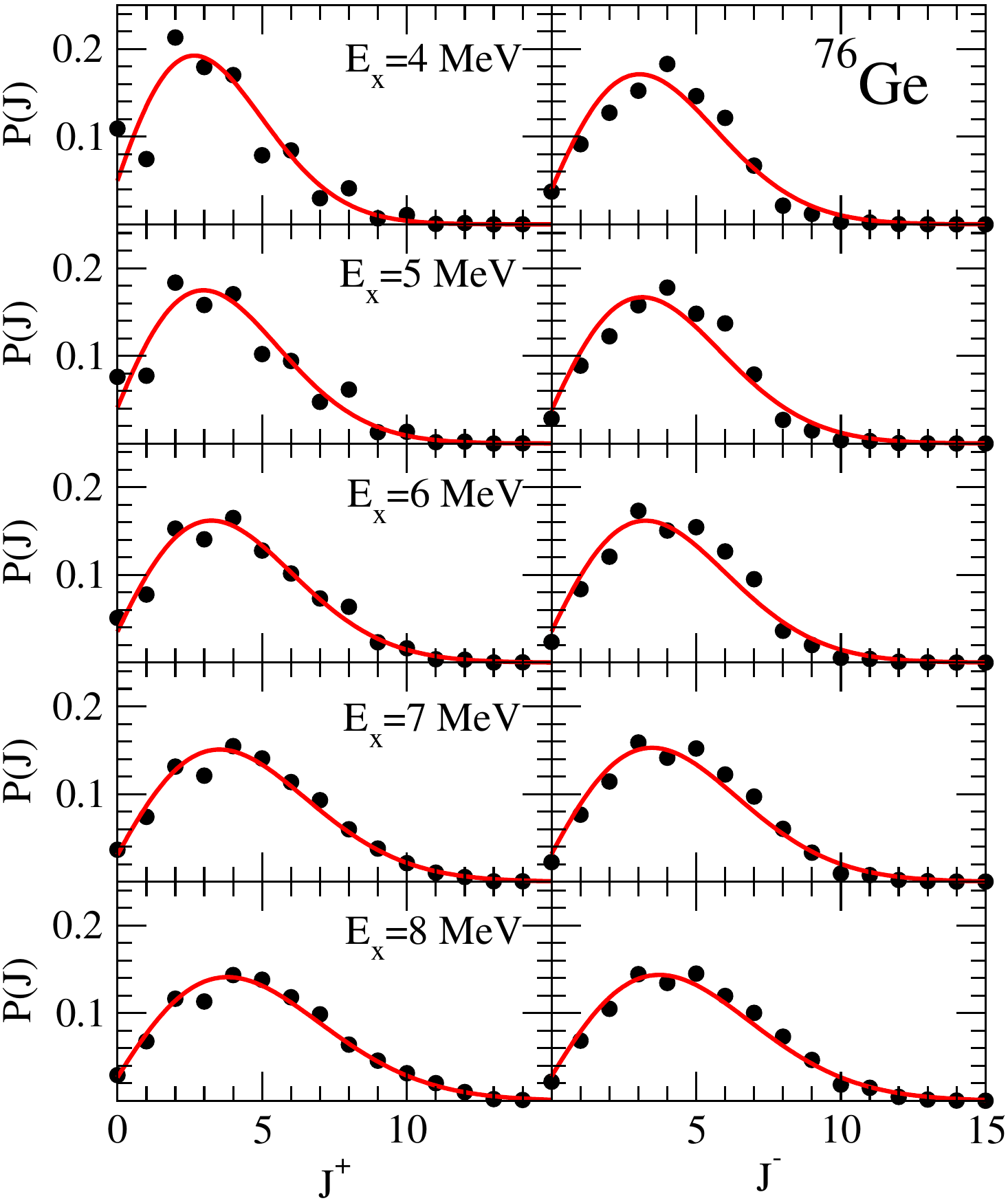}}
\caption{Angular momenta probabilities for $^{76}$Ge are shown for five excitation energies for positive- and negative-parity states. The red lines show the results obtained from Eq.~(\ref{eq:PJ}) with the spin cutoff parameter shown in Figure~\ref{fig:Ge-spincut}.}
\label{fig:Ge76-Jsq}
\end{figure}

The probability distribution of angular momenta for the three nuclei studied are shown in Figures~\ref{fig:Fe57-Jsq} - \ref{fig:Ge76-Jsq} at five distinct excitation energies. The black points are the probability distributions from the extrapolated Lanczos method, while the red lines represent the results from Eq.~(\ref{eq:PJ}) using the spin cutoff partameters computed at each excitation energy as shown in Figures~\ref{fig:Fe57-spincut} and \ref{fig:Ge-spincut}. Overall, the computed angular momenta distributions are in excellent agreement with the Bethe ansatz of Eq.~(\ref{eq:PJ}).

\section{Conclusions}

We have discussed the application of the Lanczos method to the calculation of level densities. We showed that for a given $J$ value, the $\alpha_{1-4}$ and $\beta_{1-4}$ components of the Lanczos matrix obtained from the first four Lanczos iterations provide the information sufficient to obtain the lowest eight moments of the Hamiltonian with an accuracy of approximately 1\%. We derive exact but complex equations that relate these $\alpha$ and $\beta$ matrix elements to the moments. We compare the results to calculations for matrix dimensions up to $10^6$ where exact results from full diagonalization can be obtained. We also show that the uncertainty of the moments decreases with increasing matrix dimension.

A method to extrapolate the Lanczos matrix (ELM) to the full space was presented that made use of the first eight moments of the Hamiltonian. Level densities were obtained with the ELM method and compared to exact shell-model results where possible. The ELM procedure was shown to provide an excellent representation of the asymptotic (high-energy) behavior of the level density, and with a sufficient number of actual Lanczos iterations, the ELM method was shown to provide excellent agreement with the exact shell-model level density. In some cases, a discontinuity exists between the exact Lanczos iterations and the modeled matrix elements that causes the ELM procedure to miscalculate the level density at low excitation energies. A procedure to analytically continue the the level density from the high-energy region to the lowest energy in the configuration space ($E_0$) was presented. A calculated uncertainly of about 100(10) keV in the ground-state energy is enough to obtain the level density above  with an accuracy of approximately 10(1)\% for a given model space and Hamiltonian. The calculation of the ground-state energy to within 100 keV requires on the order of 20 Lanczos iterations.

We compare the results of the ELM method with those obtained with the binomial approximation that makes use of the first four moments. In some cases with the  moments close to the Gaussian limit, the two methods give similar results. But there are other cases with the binomial method cannot be used. Finally, we compared calculations for the level density with ELM for $^{57}$Fe and $^{74,76}$Ge nuclei with those extracted from experiment. In addition, we computed $\ell = 0$ and 1 resonance spacing, $D_0$ and $D_1$, for Fe and Ge isotopes.

\acknowledgements
We gratefully acknowledges several useful discussions with S. Grimes, A. C. Larsen, Z. Meissel, A. Voinov, and K. Wendt. This work was performed under the auspices of the U.S. Department of Energy by Lawrence Livermore National Laboratory under Contract DE-AC52-07NA27344 and NSF grant PHY-1811855.

\appendix
\section{Solution for $a$- and $b$-parameters}
\label{app:A}
Given the set of moments $H_1$, $M_2$, and $R_{3-8}$, the strategy is then to find an optimal set of coefficients $ a_i$ and $ b_i$ that reproduce these moments. From Eqs.(\ref{eq:ah1})-(\ref{eq:am5}), it is clear that the moments are highly non-linear functions of the parameters $a_i$ and $b_i$. However, in general, the dominant parameters will be $a_0$ and $b_1$. For example, in the limit of a Gaussian, the odd moments are zero, and $M_2 = -2b_1$. Thus, one strategy to find the parameters is to assume that $a_{1-3}$ and $b_{2-4}$ are small and that the moments can be linearized relative to small changes in the parameters. We start with all $a_{i>0} = 0$ and solve for $b_1$ and $b_2$ using $M_2$ and $M_4$. Note that $b_2$ can be isolated with the ratio $M_4/M_2^2$, yielding a quadratic equation with two solutions, with the smallest being the most realistic. With $b_2$ found, we then use $M_2$ to fix $b_1$. Initial estimates for $a_1$ and $a_2$ can then be found from the odd moments $M_3$ and $M_5$ by truncating the analytic expressions to the leading linear terms in $a_1$ and $a_2$, yielding two coupled linear equations:
\begin{align}
M_3 \approx & 6b_1 \Big[a_1(1-4b_2)  -4a_2(1-5b_2) \Bigr] \\
M_5 \approx & 120b_1^2 \Bigl[ a_1(3+24b_2^2) - a_2(9+168b_2^2)\Bigr]
\end{align}
With these initial estimates, we then perform a Taylor expansion for the moments and truncate to first order. Representing the parameters $a_i$ and $b_i$ with the combined parameters, $p_i$, and using vector notation $\vec p = \{\vec a,\vec b\}$, a set of coupled linearized expressions for the moments can be obtained, i.e., 
\begin{equation}
M_k - M_k(\vec p) = \sum_i D_{ki}\Delta p_i,
\end{equation}
where $M_k$ is the moment for shell-model Hamiltonian and $M_k(\vec p)$ is the modeled moment evaluated from Eqs.~(\ref{eq:hh1})-(\ref{eq:mm8})  using the modeled Lanczos matrix elements $\alpha_i$ and $\beta_i$ from Eqs.~(\ref{eq:model_alpha}) and (\ref{eq:model_beta}). $D_{ki} = \frac{\partial M_k}{\partial p_i}$ is the derivative of the $k^{th}$ moment with respect to parameter $p_i$. Under the conditions that the non-linear terms are small, one can iteratively obtain the optimal parameters $\vec p$ by solving for the shift $\Delta \vec p$ and updating the derivative matrix after each iteration. In order to minimize potential effects of non-linear terms, at each iteration a fraction of the shift is taken to update the new values. In practice, half the new value was chosen, and the procedure typically finds optimal solutions in  approximately 20 iterations. 

\section{Finding the Matching Energy for the ELM}
\label{app:B}
Finding the matching energy $E_m$ to analytically continue the ELM calculation of the level density is complicated by local fluctuations in the level density due to the discrete nature of the spectrum. Thus, it is necessary to introduce a smoothing procedure in order to make use of Eqs.~(\ref{eq:temp}) and (\ref{eq:Ematch}). In this work, we made use of a low-pass filter, or Savitsky-Golay filter~\cite{savitsky-golay}, to both smooth and compute the derivative of the level density. To first order, the Savitsky-Golay filter is essentially a least-squares fit of polynomial of order $M$ to the data of interest over a region of data extending $n_L$ and $n_R$ points to the left and right of the data point of interest respectively. Here, satisfactory results were obtained by smoothing the level density directly (not the logarithm) with $M=4$ over the interval defined by $n_L=n_R=10$.


\begin{thebibliography}{9}
\bibitem{r-process} M. Burbidge, G. Burbidge, W. Fowler, and F. Hoyle, Rev. Mod. 
     Phys. 29, 547 (1957). 
\bibitem{merg} D. Kasen, B. Metzger, J. Barnes, E. Quataert,  and E. 
Ramirez-Ruiz, 
     Nature Vol. {\bf 551}, 80 (2017). 
\bibitem{surr} J. E. Escher, J. T. Burke, F. S. Dietrich, N. D. Scielzo, 
I. J. Thompson, W. Younes, 
     Rev. Mod. Phys. {\bf 84}, 353 (2012). 
\bibitem{Hauser-Feshbach} W. Hauser, H. Feshbach, Phys. Rev. {\bf 87}, 366 
     (1952). 
\bibitem{Goriely} S. Goriely, M. Samyn, J.M. Pearson, Phys. Rev. C {\bf 75} 
     (2007) 064312; S. Goriely, S. Hilaire, A. J. Koning, Phys. Rev. C {\bf 78} 
     (2008) 064307. 
\bibitem{AFMC} C. W. Johnson, S. E. Koonin, G. H. Lang, W. E. Ormand, 
Phys. Rev. 
     Lett {\bf 69}, 3157 (1992). 
\bibitem{AFMC-2} G. H. Lang, C. W. Johnson, S. E. Koonin, and W. E. Ormand, Phys. 
     Rev. C {\bf 48}, 1518 (1993) 
\bibitem{AFMC-rho} S. E. Koonin, D. J. Dean, and K. Langanke, Phys. Rep. {\bf 
     278}, 1 (1997); H. Nakada and Y. Alhassid, Phys. Rev. Lett. {\bf 79}, 2939 
     (1997); W. E. Ormand, Phys. Rev. C {\bf 56}, R1678 (1997); Y. Alhassid, S. Liu, 
     and H. Nakada, Phys. Rev. Lett. {\bf 83}, 4265 (1999); H. Nakada and Y. 
     Alhassid, Phys. Rev. C{\bf 78}, 051304(R) (2008); C. \"Ozen, Y. Alhassid, and H. 
     Nakada, Phys. Rev. C {\bf 91}, 034329 (2015).
\bibitem{Mon75} K.~K.~Mon and J.~B.~French, Ann. of Phys. {\bf 95},1 (1975); 
S.~S.~M.~Wong and J.~.B.~French, Nucl. Phys. {\bf A198}, 188 (1972); J.~B.~French and V.~K.~B.~Kota, Ann. Rev. Nucl. Part. Sci. {\bf 32}, 35 (1982); J.~B.~French and V.~K.~B.~Kota, Phys. Rev. Lett. {\bf 51}, 2183 (1983);  Z.~Pluhar and H.~A.~Weidenmuller, Phys. Rev. C {\bf 38}, 1046 (1988); A.~.P.~Zuker, Phys. Rev. C {\bf 64}, 021303(R) (2001).
\bibitem{mom} R. A. Sen'kov and M. Horoi, Phys. Rev. C {\bf 82}, 024304 
(2010)
\bibitem{Horoi}R. A. Sen'kov, M. Horoi, and V. G. Zelivinsky, Comp. Phys. Comm. {\bf 184}, 215 (2013).
\bibitem{shi} N. Shimizu, Y. Utsuno, Y. Futamura, T. Sakurai, T. Mizusaki, 
     and T. Otsuka, Phys. Lett. B {\bf 753}, 13 (2016) 
\bibitem{26} B. Jegerlehner, arXiv:hep-lat/9612014, (1996). 
\bibitem{27} S. Yamamoto, T. Sogabe, T. Hoshi, S.-L. Zhang, T. Fujiwara, 
    J. Phys. Soc. Jpn. {\bf 77}, 114713 (2008). 
\bibitem{shell-model} R. D. Lawson, {\it Theory of the Nuclear Shell Model}, 
     (Clarendon, Oxford, 1980); P. J. Brussaard and P. W. M. Glaudemans, {\it Shell 
     Model Applications in Nuclear Spectroscopy}, (North-Holland, Amsterdam, 1977). 
\bibitem{Lanczos} C. Lanczos, J. Res. Nat. Bur. Stand. {\bf 45}, 252 (1950); J. 
     H. Wilkinson, {\it The Algebraic Eigenvalue Problem}, (Clarendon, Oxford, 1965);
     R.~R. Whitehead, A.~Watt, B.~J.~Cole, and I.~Morrison, Adv, in Nucl. Phys. {\bf 9}, 123 (1977).
\bibitem{zuker}  A.P. Zuker, L. Waha Ndeuna, F. Nowacki, and E. Caurier, Phys. 
    Rev. C {\bf 64}, 021394(R) (2001); E. Caurier, G. Martinez-Pinedo, F. Nowacki, A. Poves, 
    A.P. Zuker, rev. Mod. Phys. {\bf 77}, 427 (2005).
\bibitem{nushellx} B. A. Brown and W. D. M. Rae, 
     Nuclear Data Sheets {\bf 120}, 115 (2014). 
\bibitem{gxpf1a} M. Honma, T. Otsuka, B.A. Brown and T. Mizusaki, 
     Euro. Phys. Jour. A {\bf 25} Suppl. 1, 499 (2005). 
\bibitem{Muk} S. Mukhopadhyay, B. P. Crider, B. A. Brown, S. F. Ashley, A. Chakraborty, A. Kumar, M. T. McEllistrem, E. E. Peters, F. M. Prados-Est{\' e}vez, and S. W. Yates, Phys. Rev. C {\bf 95}, 014327 (2017).
\bibitem{Homna_2009} M. Honma, T. Otsuka, T. Mizusaki, and M. Hjorth-Jensen, 
     Phys. Rev. C {\bf 80}, 064323 (2009). 
\bibitem{ormand} W. E. Ormand, Int. J. Mod. Phys. {\bf 14}, 67 (2005).
\bibitem{Gilbert-Cameron} A. Gilbert and A. G. W. Cameron, Can. J. Phys. 43, 1446 
     (1965). 
\bibitem{CERN} B. A. Brown and W. E. Ormand, CERN Proc. 1, 21 (2019). DOI: 10.23727/CERN-Proceedings-2019-001
\bibitem{zuker_2} A.~P.~ Zuker, Phys. Rev. C {\bf 64}, 021303 (2001).
\bibitem{mug} Atlas of Neutron Resonances 
     Volume 1: Resonance Properties and Thermal Cross Sections Z= 1-60 
     by Said F. Mughabghab, Elsevier (2018) 
     (doi.org/10.1016/B978-0-44-463769-7.00001-4). 
\bibitem{nndc} Data from the NNDC On-Line Data Service database as of 
August 2017 
     http://nndc.bnl.gov/nudat2/ 
\bibitem{Algin} E. Algin, {\it et al.}, Phys. Atomic Nuclei {\bf 70}, 1634 (2007).
\bibitem{Voinov}A. Voinov, {\it et al.}, Phys. Rev. Lett. {\bf 93}, 142504 (2004).
\bibitem{Lar17} A. C. Larsen {\it et al.}, J. Phys. G {\bf 44}, 065005 (2017); 
     and private communication (2018). 
\bibitem{Voinov-2} A. V. Voinov, {\it et al.}, Phys. Rev. C {\bf 99}, 054609 (2019).
\bibitem{Ge74-Oslo} T. Renstr\o m, H.-T. Nyhus, H. Utsunomiya, R. Schwengner, S. Goriely, A. C. Larsen, D. M. Filipescu, I. Gheorghe, L. A. Bernstein, D. L. Bleuel, T. Glodariu, A. Görgen, M.Guttormsen, T. W. Hagen, B. V. Kheswa, Y.-W. Lui, D. Negi,I. E. Ruud, T. Shima, S. Siem, K. Takahisa, O. Tesileanu, T. G.Tornyi, G. M. Tveten, and M. Wiedeking, Phys. Rev. C93, 064302 (2016).
\bibitem{Ge76-Oslo} A. Spyrou, S. N. Liddick, A. C. Larsen, M. Guttormsen, K. Cooper, A. C. Dombos, D. J. Morrissey, F. Naqvi, G. Perdikakis, S. J. Quinn, T. Renstrom, J. A. Rodriguez, A. Simon, C. S. Sumithrarachchi, and R. G. T. Zegers, Phys. Rev. Lett. {\bf 113}, 232502 (2014).
\bibitem{bethe} H. A. Bethe, Phys. Rev. {\bf 50}, 332 (1936); 
     Rev. Mod. Phys. {\bf 9}, 69 (1937). 
\bibitem{savitsky-golay} A. Savitsky and M. J. E. Golay, Anal. Chem. {\bf 36}, 1627 (1964); R. W. Hamming, {\it Digital Filters}, 2nd ed. (Englewood Cliffs, NJ: Prentice Hall, 1983); M. U. A. Bromba, Anal. Chem. {\bf 53}, 1583 (1981); W. H. Press, S. A. Teukolsky, W. T. vettering, and B. P. Flannery, {\it Numerical Recipes in Fortran}, 2nd ed. (Cambridge University Press, New York,1992), p. 644-649.
\end{thebibliography}
\end{document}